\RequirePackage{lineno}
\documentclass[letterpaper,twocolumn,prl,aps,superscriptaddress,amsmath,amssymb,floatfix]{revtex4}
\usepackage{mathptmx}
\usepackage[utf8]{inputenc}
\usepackage{color}
\usepackage{amsmath}
\usepackage{amssymb}
\usepackage{graphicx}
\usepackage{esint}
\usepackage[unicode=true,
 bookmarks=true,bookmarksnumbered=false,bookmarksopen=false,
 breaklinks=false,pdfborder={0 0 1},backref=false,colorlinks=true]
 {hyperref}
\hypersetup{
 linkcolor=magenta,urlcolor=blue,citecolor=blue,pdfstartview={FitH},hyperfootnotes=false}

\makeatletter

\DeclareFontEncoding{LGR}{}{}

\ProvideTextCommand{\~}{LGR}[1]{\char126#1}

\newcommand{\lyxmathsym}[1]{\ifmmode\begingroup\def\b@ld{bold}
  \text{\ifx\math@version\b@ld\bfseries\fi#1}\endgroup\else#1\fi}

\usepackage{textcomp}
\usepackage{epstopdf}

\usepackage{amsfonts}

\pdfpageheight\paperheight
\pdfpagewidth\paperwidth

\@ifundefined{textcolor}{}{%
 \definecolor{BLACK}{gray}{0}
 \definecolor{WHITE}{gray}{1}
 \definecolor{RED}{rgb}{1,0,0}
 \definecolor{GREEN}{rgb}{0,1,0}
 \definecolor{BLUE}{rgb}{0,0,1}
 \definecolor{CYAN}{cmyk}{1,0,0,0}
 \definecolor{MAGENTA}{cmyk}{0,1,0,0}
 \definecolor{YELLOW}{cmyk}{0,0,1,0}
}

\usepackage{xcolor}\usepackage{soul}
\definecolor{blue}{rgb}{0,0,1}
\definecolor{red}{rgb}{1,0,0}
\definecolor{green}{rgb}{0,1,0}

\makeatother

\begin{document}

%\setpagewiselinenumbers
%\modulolinenumbers[5]
%\linenumbers

%\title{Analyzing the dephasing of optically trapped neutral atom: a full quantum approach}
\title{Gate fidelity, dephasing, and “magic” trapping of optically trapped neutral atom}
\affiliation{State Key Laboratory of Quantum Optics and Quantum Optics Devices, and
Institute of Opto-Electronics, Shanxi University, Taiyuan 030006,
China}
\affiliation{Collaborative Innovation Center of Extreme Optics, Shanxi University,
Taiyuan 030006, China}

\author{Pengfei Yang}

\affiliation{State Key Laboratory of Quantum Optics and Quantum Optics Devices, and
Institute of Opto-Electronics, Shanxi University, Taiyuan 030006,
China}
\affiliation{Collaborative Innovation Center of Extreme Optics, Shanxi University,
Taiyuan 030006, China}

\author{Gang Li}
\email{gangli@sxu.edu.cn}

\affiliation{State Key Laboratory of Quantum Optics and Quantum Optics Devices, and
Institute of Opto-Electronics, Shanxi University, Taiyuan 030006,
China}
\affiliation{Collaborative Innovation Center of Extreme Optics, Shanxi University,
Taiyuan 030006, China}

\author{Zhihui Wang}

\affiliation{State Key Laboratory of Quantum Optics and Quantum Optics Devices, and
Institute of Opto-Electronics, Shanxi University, Taiyuan 030006,
China}
\affiliation{Collaborative Innovation Center of Extreme Optics, Shanxi University,
Taiyuan 030006, China}

\author{Pengfei Zhang}

\affiliation{State Key Laboratory of Quantum Optics and Quantum Optics Devices, and
Institute of Opto-Electronics, Shanxi University, Taiyuan 030006,
China}
\affiliation{Collaborative Innovation Center of Extreme Optics, Shanxi University,
Taiyuan 030006, China}

\author{Tiancai Zhang}
\email{tczhang@sxu.edu.cn}
\affiliation{State Key Laboratory of Quantum Optics and Quantum Optics Devices, and
Institute of Opto-Electronics, Shanxi University, Taiyuan 030006,
China}
\affiliation{Collaborative Innovation Center of Extreme Optics, Shanxi University,
Taiyuan 030006, China}

%\date{\today}

\begin{abstract}
The fidelity of the gate operation and the coherence time of neutral atoms trapped in an optical dipole trap are figures of merit for the applications. The motion of the trapped atom is one of the key factors which influence the gate fidelity and coherence time. The motion has been considered as a classical oscillator in analyzing the influence. Here we treat the motion of the atom as a quantum oscillator. The population on the vibrational states of the atom are considered in analyzing the gate fidelity and decoherence. We show that the fidelity of a coherent rotation gate is dramatically limited by the temperature of a thermally trapped atom. We also show that the dephasing between the two hyperfine states due to the thermal motion of the atom could rephase naturally if the differential frequency shift is stable and the vibrational states do not change. The decoherence due to the fluctuations of the trap laser intensity is also discussed. Both the gate fidelity and coherence time can be dramatically enhanced by cooling the atom into vibrational ground states and/or by using a blue-detuned trap. More importantly, we propose a ``magic'' trapping condition by preparing the atom into specific vibrational states.

\end{abstract}

\maketitle
\section{Introduction}
Neutral atoms, with long-lived internal electronic states, trapped in optical dipole trap (ODT) are one of the basic systems for quantum metrologies~\cite{Jun2008,Derevianko2011}, quantum simulations~\cite{Bloch2008,Labuhn2016,Browaeys2020}, and quantum information processing~\cite{Saffman2010,Reiserer2015}. The fidelity of a coherent rotation gate and the coherence time between two fiducial states are figures of merit for these applications. The fiducial states are usually chosen from the Zeeman states in the ground hyperfine levels. Both the gate fidelity and the coherence time are assumed to be limited by the variance of the differential frequency shift (DFS) due to the motion of the atom and the noise of the trap laser. For the optically trapped atoms, the energies of these sublevels are subject to the fluctuations of the trap beam and the surrounding magnetic fields. The gate operation will be deteriorated due to the variance of the detuning between the driving field and the atomic transition. The evolution of the states will dephase to each other due to the resulting fluctuations on the energy levels. In order to suppressing the DFS, a series of ``magic'' trapping conditions, where the differential energy shift between the fiducial states is immune to the fluctuations, are proposed and experimentally tested~\cite{Lundblad2010,Dudin2010,Kim2013,Kazakov2015,Yang2016, Flambaum2008,Derevianko2010,Chicireanu2011,Carr2016,Li2019}. The infidelity of the gate operation can be suppressed from 0.01~\cite{Xia2015} to $0.5\times 10^{-5}$ \cite{Sheng2018}. Over one second of coherence time $T_2$ has also been realized either in the red-detuned trap~\cite{Guo2020} or the blue-detuned trap~\cite{Wu2019, Li2020}.

In the preceding works~\cite{Kuhr2005, Rosenfeld2011, Gerasimov2021} on analyzing the DLS and the corresponding dephasing mechanics, people usually treated the motion of the atom inside the ODT as a classical harmonic oscillator. The inhomogeneous and homogeneous dephasing factors are classified. The inhomogeneous dephasing is mainly caused by the differential light shift (DLS) associated with the motion of the atom and can be recovered by the spin-echo technique~\cite{Hahn1950,Andersen2003,Kuhr2003}. The homogeneous dephasing is from the fluctuations of DLS induced by the noises on magnetic field and ODT beam, e.g., the noises on the power, the frequency, and the pointing direction etc. However, we know the motion of the atom in an ODT is actually a quantum oscillator and the atom occupies a serial of vibrational quantum states depending on the temperature. In this sense, the DFS depends not only on the trap potential itself but also on the vibrational quantum states for a given atomic distribution on the vibrational states. Thus, the former views should be reconsidered. 

\begin{figure}
\includegraphics[width=0.75 \columnwidth]{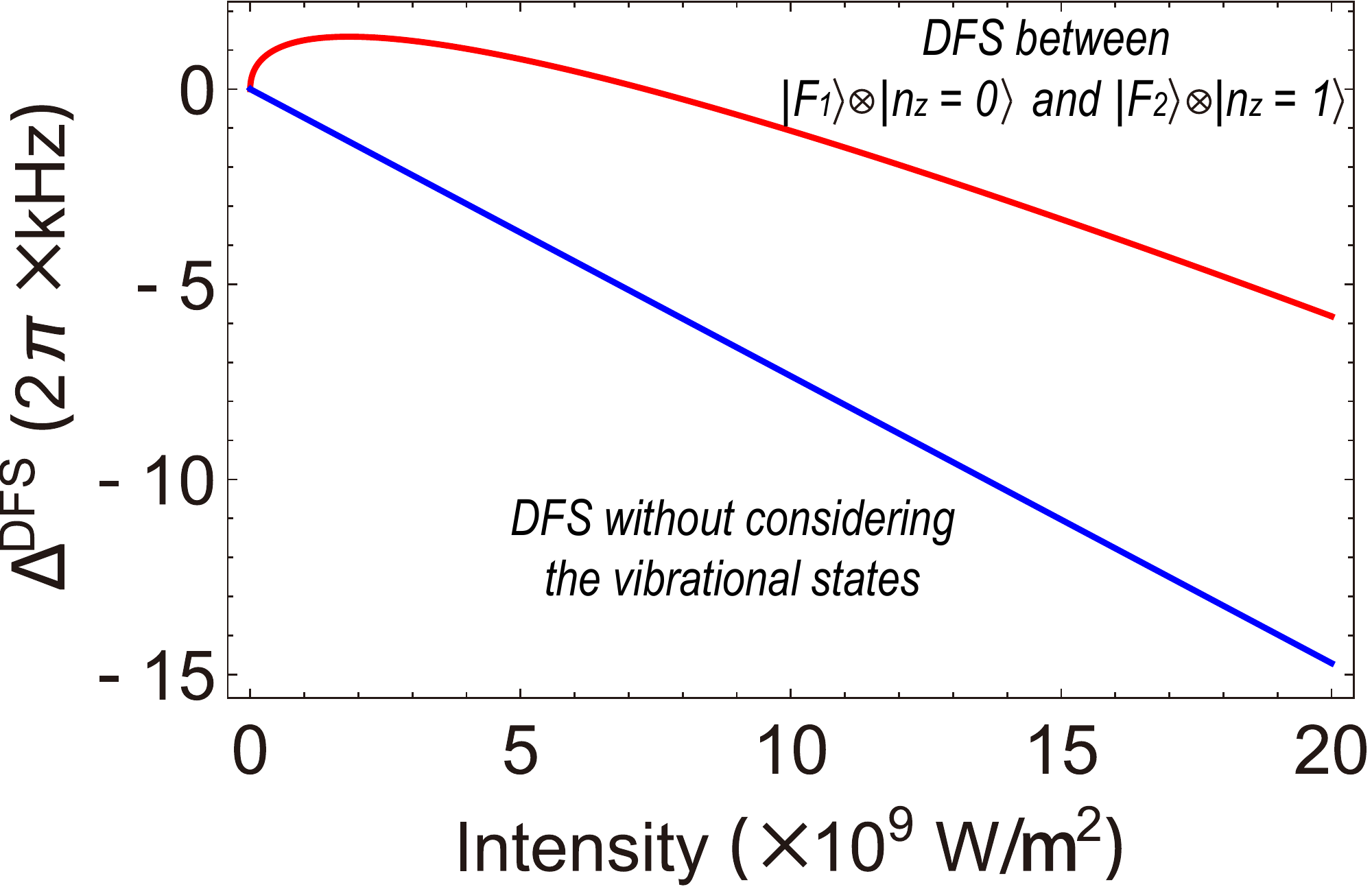}
\caption{\label{fig0} The comparison of DFS versus the light intensity at the trap bottom with and without considering the vibrational quantum states for a cesium atom trapped in a red-detuned ODT. The parameters used for the simulation: trap wavelength $\lambda_T=1064$ nm, trap size (beam waist radius) $w_0=2.1\ \mu\text{m}$, $\eta=1.68 \times 10^{-4}$ for cesium atom.}
\end{figure}

Here in this paper we provide theoretical analysis on the operating errors of the coherent $\pi/2$ rotation gate and the dephasings of an atom in ODT in the context of a quantum oscillator. In the new picture, the overall DFS will contain a new term which is directly connected to the vibrational quantum number. As shown in Fig. \ref{fig0}, the DFS with the atom prepared in specific vibrational quantum states shows a nonlinear dependence on the trap intensity, which is totally different from the linear dependence without considering the vibrational states. The new term will bring several new findings on the gate fidelity and the coherence time. 1) The gate fidelity is dramatically influenced by the vibrational state distribution (the temperature) of the atom. Thus, the fidelity can be enhanced by squeezing the state distribution, e.g., cooling the atom to a lower temperature. 2) Remarkably, we found that the thermal motion of the atom (Bose distribution on the vibrational states) would not make the fiducial states lose their phase during the free evolution. The fringe visibility of the Ramsey interference drops in short time scale due to the overlap of a series of interfering signals with different frequencies, which are determined by the DLS associated with the vibrational states. The fringe will naturally recover as long as the DLS is stable and the vibrational states do not change. However, the fluctuation of DLS and the heating of the atom is inevitable in real system, the recovery of the fringe will be inhibited. Consequently, the fringe will only be recovered by the spin-echo process in a relatively shorter time. 3) The dephasing due to the fluctuation of the DLS, which is mainly induced by the trap intensity noise, in both the red-detuned and blue-detuned ODTs are revisited. Because of the intrinsic advantages of small DFS and low parametric heating rate in blue-detuned ODT, the longer coherence time will be expected. 4) The new term in the DLS also inspire a new kind of ``magic'' trapping condition by prepare the atom in specific vibrational quantum states. We discussed a series of ``magic'' conditions in both the red-detuned and blue-detuned ODTs.

The remaining part of the paper is organized as follows. In Sec.II the DFS between two hyperfine ground states of trapped atom in a red-detuned optical trap is revisited by treating the trapped atom as a quantum oscillator. The fidelity of coherent rotation gate is analyzed with a trapped thermal atom in Set.III. Then, the dephasing processes of a trapped thermal atom are discussed in Sec.IV. Next, in Sec.V, a new ``magic'' trapping condition is presented. Finally, a conclusion is given in Sec.VI.

\begin{figure*}
\includegraphics[width=17 cm]{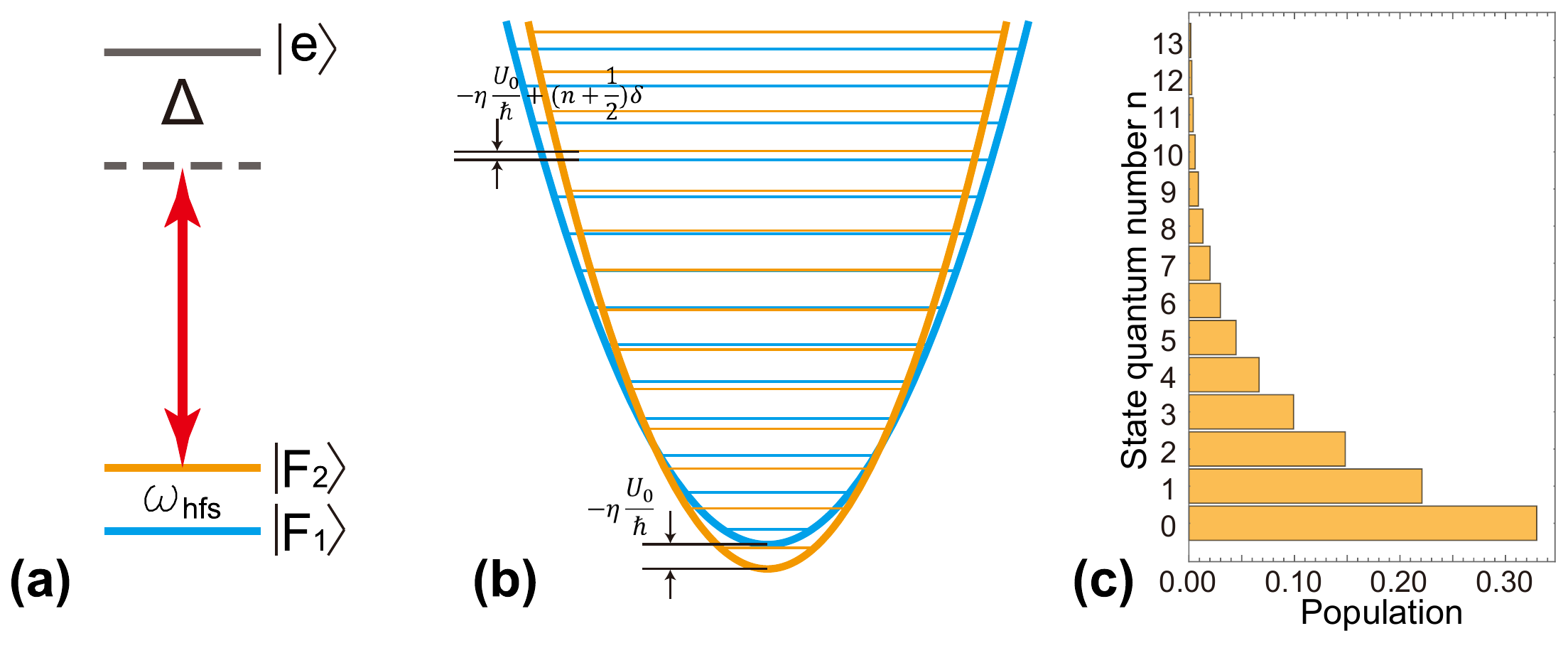}
\caption{\label{fig1} A conceptual energy level scheme of the interaction between atom with two ground hyperfine states and the light field of an ODT. (b) The comparison of the vibrational quantum states for the atom in two ground states at the bottom of the ODT. Due to the difference of one-photon detuning there is energy frequency shifts $(n+1/2) \delta$ with $\delta$ the differential oscillation frequency shift for vibrational state $n$. (c)  An example of population distribution for thermal atom on vibrational state. }
\end{figure*} 

\section{DLS distribution of an atom in ODT}

An ODT is formed by the spatial-dependent light shift of ground state when the atom interacts with a far-off-resonant laser beam. For a two-level atom, the light shift of the ground state reads~\cite{Grimm2000}
\begin{equation}
\Delta E=\frac{3 \pi c^2}{2 \omega_0^3} \frac{\Gamma}{\Delta} I,\label{eq1}
\end{equation}
where $c$ is the velocity of light; $\omega_0$ is the resonant frequency of the atom transition; $\Gamma$ is the decay rate of the excited state; $\Delta$ is the detuning of the light frequency to the atomic transition; and $I$ is the intensity of the laser beam. A trap is formed when there exists maxima or minima in the spatial distribution of the laser intensity. The sign of the trapping potential depends on the detuning $\Delta$. Thus, the atom populated on the electronic ground state can be trapped locally in the light field.

Usually, two electronic ground states in different hyperfine levels are adopted as the fiducial states for the applications. The light shifts for the two ground states are unequal to each other due to the additional hyperfine splitting $\omega_\text{hps}$ between the two states. According to Eq.~(\ref{eq1}), the different frequency detuning between the corresponding atomic transitions and the trapping laser field will give a differential DLS between the two ground states. Then, the phase between the two electronic ground states will be disturbed when the DLS is fluctuating because of the movement of trapped thermal atom (inhomogeneous dephasing factor) in the trap and the intensity noise of the laser beam (homogeneous dephasing factor). These dephasing processes has been analyzed as the atom motion being treated classically~\cite{Kuhr2003,Kuhr2005}. However, this treatment is not appropriate in a typical used ODT for single atoms, where a small trap volume is pursued in order to enhance the light-assisted two-body loss rate~\cite{Schlosser2002}. The atom in the ODT behaves actually as a quantum oscillator. When the trapped atom is cooled by optical molasses, the energy follows a thermal Boltzmann distribution and the atom occupies a series of separate vibrational states. 

We first consider a typical trap by focusing a red-detuned Gaussian laser beam. So, the trap potential is negative and the atom is trapped in the potential minima where the laser intensity is a maximum. As shown in Fig. \ref{fig1}(a), the two fiducial states are $|F_{1}\rangle$ and $|F_{2}\rangle$, which are two Zeeman sublevels belong to two hyperfine states, and $\omega_\text{hfs}$ is the hyperfine frequency splitting. The corresponding potential depths induced by a far-detuned ODT laser beam, which has one-photon detuning $\Delta$ to atomic transition $|F_2\rangle \leftrightarrow |\text{e}\rangle$, are $U_{1}=\left|\Delta E_{1}^\text{max} \right|$ and $U_{2}=\left|\Delta E_{2}^\text{max} \right|$, respectively. $\Delta E_{1(2)}^\text{max}$ is the maximum light shift for the electronic ground state. By using Eq.~(\ref{eq1}) we have $\frac{U_1}{U_2}=\frac{\Delta}{\Delta-\omega_\text{hfs}}$. Supposing the waist radius of the trapping beam is $w_0$, the Rayleigh length on the laser propagating direction ($z$-axis) is then $L_f=\frac{\pi w_0^2}{\lambda}$ with $\lambda$ the laser wavelength. The temperature of the trapped atom is usually much lower than the trap depth. Therefore, the trap potentials can be approximated by parabolic functions. The atom behaves as a three-dimensional (3D) quantum oscillator, and the oscillation frequencies on transverse and longitudinal directions are read as \cite{Saffman2005}
\begin{equation}
\omega_{r,1(2)}=\omega_{x(y),1(2)}=\frac{2}{w_0} \sqrt{\frac{U_{1(2)}}{m}} \label{eq2}
\end{equation}
and
\begin{equation}
\omega_{z,1(2)}=\frac{\sqrt{2}}{L_f} \sqrt{\frac{U_{1(2)}}{m}},\label{eq3}
\end{equation}
where $m$ is the mass of the trapped atom. The difference between two oscillation frequencies on one axis with atom in states $|F_{1}\rangle$ and $|F_{2}\rangle$ is then 
\begin{equation}
\delta_{q}=\omega_{q,2}-\omega_{q,1}, \label{eq4}
\end{equation} 
where $q=x,\ y,\ \text{or } z.$

For simplicity, we first deal with a one-dimensional trap. Thus, the subscripts representing the axis number in Eqs. (\ref{eq2})-(\ref{eq4}) can be omitted. The DFS between two electronic states with the atom on the same vibrational state $|n\rangle$ is then
\begin{equation}
\Delta^\text{DFS}_n=-\left( U_2-U_1 \right) / \hbar+(n+\frac{1}{2}) \delta. \label{eq5}
\end{equation}
The first term on the rhs (right hand side) is the DLS induced by the trap depth and the second term is the additional shifts due to the unequal vibrational frequencies, which depends on the vibrational quantum number $n$. In a far-off-resonant trap (FORT), $\Delta \gg \omega_\text{hfs}$, we define $U_0=U_1 \approx U_2$. Equation (\ref{eq5}) can be approximated by
\begin{equation}
\Delta^\text{DFS}_n=-\eta \frac{U_0}{\hbar}+(n+\frac{1}{2}) \delta \label{eq6}
\end{equation}
with $\eta=\left| \frac{\omega_\text{hfs}}{\Delta} \right|$. Figure \ref{fig1}(b) gives a conceptual drawing to explain the DFS induced by the difference in oscillation frequencies when the atom is in different fiducial state.

In a three-dimensional (3D) trap the DFS between the two atomic states is then
\begin{equation}
\Delta^\text{DFS}=-\eta \frac{U_0}{\hbar}+\sum_{q=x,y,z} (n_q+\frac{1}{2}) \delta_q, \label{eq7}
\end{equation}
where $q$ represents the three-oscillation axis of the trap. Thus, $n_q$ and $\delta_q$ are the vibrational quantum (phonon) number and the difference of the oscillation frequencies on the corresponding axis. For the case with a trap formed by a red-detuned laser beam, we also have $\delta_x=\delta_y=\left| \frac{\omega_\text{hfs}}{\Delta} \right| \frac{1}{w_0} \sqrt{\frac{U_0}{m}}=\frac{\eta}{2}\omega_{r,0}$ and $\delta_z=\left| \frac{\omega_\text{hfs}}{\Delta} \right| \frac{1}{\sqrt{2} L_f} \sqrt{\frac{U_0}{m}} = \frac{\eta}{2}\omega_{z,0}$ from Eqs. (\ref{eq2}) and (\ref{eq3}).

Consider an atom is directly loaded from a magneto-optical trap (MOT), where the energy of the atom follows the Boltzmann distribution. The trapped atom in the dipole trap has a thermal distribution on the vibrational states. Suppose the temperature of the trapped atom is $T$,  the average vibrational quantum number on each direction for the atom is $\langle n_q \rangle = \frac{k_\text{B}T}{2 \hbar \omega_q}$ with $k_\text{B}$ the Boltzmann's constant, $\hbar$ the Planck's constant, and $\omega_q$ the oscillation frequency on the corresponding axis. The population of the atom on the vibrational state $n$ follows the Bose distribution
\begin{equation}\label{eq8}
P_n=\frac{ \langle n_q \rangle ^n }{\left( \langle n_q \rangle +1 \right)^{(n+1)} } 
\end{equation}
on each oscillating direction.  Here we see that the population on the ground state with $n=0$ is most, and the population on higher state is gradually reduced along with the quantum number $n$ [see Fig.~\ref{fig1}(c)]. 

\section{Gate operating error}

The coherent rotation of the atomic states is usually achieved by driving the atom with a resonant microwave or Raman lasers. However, due to the thermal distribution of the atom on the vibrational states and the variances of $n_q$-dependent DFS, the driven field would not be resonant with all the atomic transitions between $|F_1\rangle \otimes |n_q\rangle$ and $|F_2\rangle \otimes |n_q\rangle$. This will cause errors in the rotating operations. We will analyze the oprating errors of a $\pi/2$-pulse (Hadamard gate) in the following.

We use the Bloch equations~\cite{Foot2004} to describe the rotation process. Suppose the rotation Rabi frequency is $\Omega$. When frequency detuning $\Delta'$ exists, the Bloch equations to describe the rotating process are
\begin{subequations}
\begin{align} 
\dot{u}&=\Delta' v, \label{eq9a} \\
\dot{v}&=-\Delta' u + \Omega w, \label{eq9b} \\
\dot{w}&=-\Omega v, \label{eq9c}
\end{align}
\end{subequations}
where $u$ and $v$ are the real and imaginary parts of the atomic density matrix; $w$ is the population difference between the two fiducial states. The Bloch equations can also be expressed by vector equation
\begin{equation}\label{eq10}
\dot{\mathbf{R}}=\mathbf{R} \times \mathbf{W},
\end{equation}
with the state vectors $\mathbf{R}=(u,v,w)^\mathbf{T}$ and driving vector $\mathbf{W}=(\Omega,0,\Delta')^\mathbf{T}$, where the superscript $\mathbf{T}$ means the transposition. The evolution of the state vector $\mathbf{R}$ can be seen as rotation around vector $\mathbf{W}$. The states $|F_1\rangle$ and $|F_2\rangle$ can be represented by vectors $(0,0,-1)^\mathbf{T}$ and $(0,0,1)^\mathbf{T}$, respectively. The application of a $\pi/2$ pulse usually represents a rotation with angle $\pi/2$ around the driving vector on the Bloch sphere. If $\Delta'=0$, the $\pi/2$-rotation can be described by a matrix
\begin{equation}\label{eq11}
\Theta_{\pi/2}=
\begin{pmatrix}
1 & 0 & 0 \\
0 & 0 & 1 \\
0 & -1 & 0
\end{pmatrix}.
\end{equation}
The state after the application of a $\pi/2$ pulse on an initial state $\mathbf{R}(t_{0})$ is then
\begin{equation}\label{eq12}
\mathbf{R}(t_{\pi/2})=\Theta_{\pi/2} \mathbf{R}(t_{0}).
\end{equation}

However, as we discussed before, the trapped thermal atom has a population on different vibrational states and population on the ground vibrational state is most. So, we assume the driven field is resonant with the atomic transitions with vibrational quantum number $n_q=0$ and the Rabi frequency is $\Omega_0$. The time duration of a $\pi/2$ pulse is defined by $t_{\pi/2}=\pi/(2 \Omega_0)$. The atomic transitions on higher vibrational states 
are then off-resonant and the frequency detuning is $\Delta'= \sum_{q=x,y,z} n_q \delta_q$. The Rabi frequencies for these transitions are then $\Omega_{n_x,n_y,n_z}=\sqrt{\Delta'^2 +\Omega_0^2}$. The application of the driven field with same duration time $t_{\pi/2}=\pi/(2 \Omega_0)$ will result a rotation angle $\theta=\frac{\pi}{2} \frac{\Omega_{n_x,n_y,n_z}}{\Omega_0}=\frac{\pi}{2} \frac{\sqrt{\Delta'^2 +\Omega_0^2}}{\Omega_0}$ on Bloch sphere around vector $\mathbf{W}=(\Omega_0,0,\Delta')^\mathbf{T}$. The rotation can be expressed by matrix
\begin{equation}\label{eq13}
\Theta'_{\pi/2}=
\begin{pmatrix}
\frac{\Omega_0^2+\Delta'^2 \cos{\theta} }{\Omega_0^2+\Delta'^2} & -\frac{\Delta' \sin{\theta} }{\sqrt{\Omega_0^2+\Delta'^2}} & \frac{\Omega_0 \Delta' (1- \cos{\theta}) }{\Omega_0^2+\Delta'^2} \\
\frac{\Delta' \sin{\theta} }{\sqrt{\Omega_0^2+\Delta'^2}} & \cos{\theta} & \frac{\Omega_0 \sin{\theta} }{\sqrt{\Omega_0^2+\Delta'^2}} \\
\frac{\Omega_0 \Delta' (1- \cos{\theta}) }{\Omega_0^2+\Delta'^2} & -\frac{\Omega_0 \sin{\theta} }{\sqrt{\Omega_0^2+\Delta'^2}} & \frac{\Omega_0^2 \cos{\theta}+\Delta'^2 }{\Omega_0^2+\Delta'^2}
\end{pmatrix}.
\end{equation}

\begin{figure}
\includegraphics[width=8.5 cm]{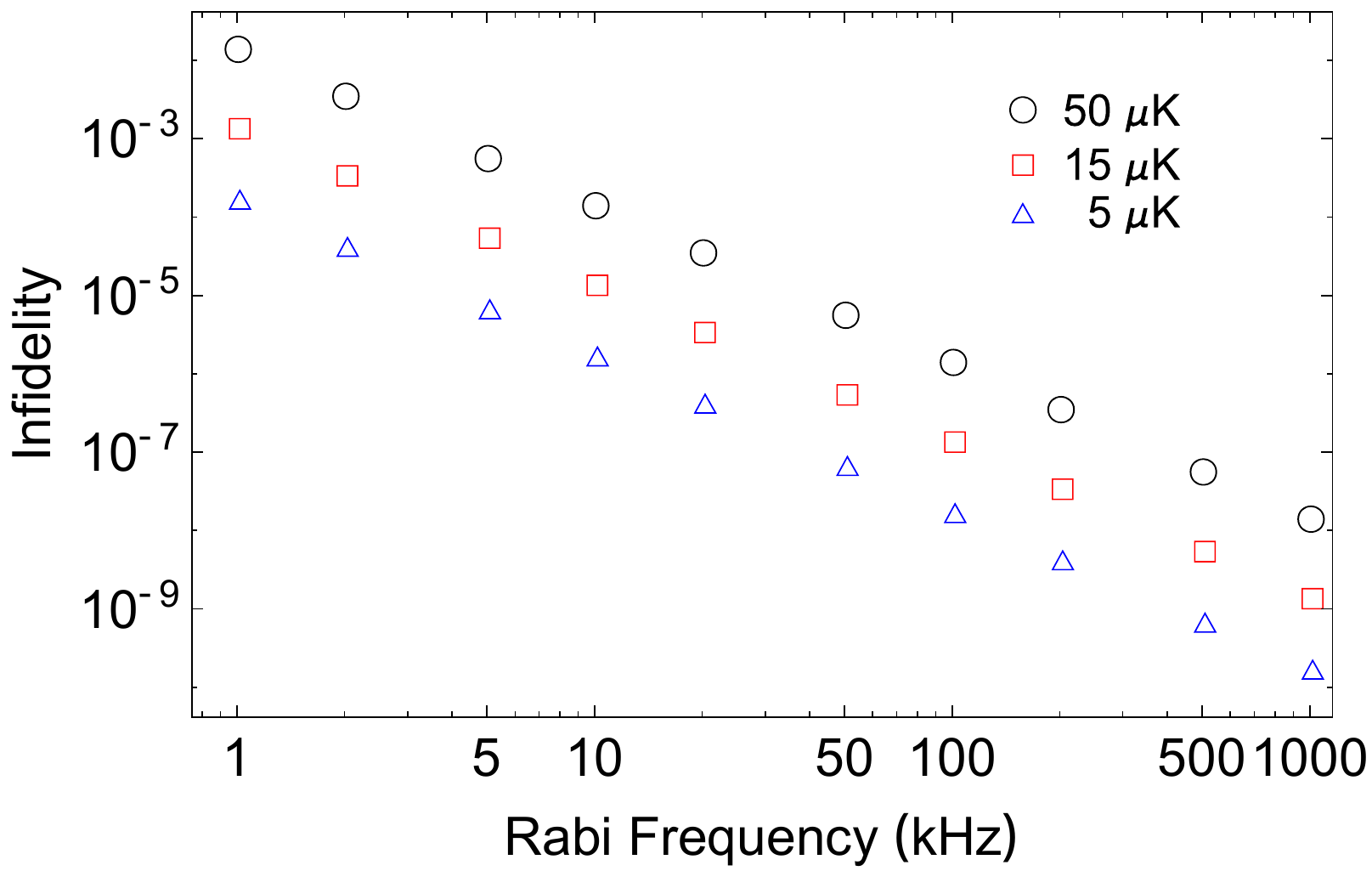}
\caption{\label{fig2} The simulation of the error (the infidelity $1-\mathbf{F}$) for a $\pi/2$ pulse between the two ground hyperfine states versus the Rabi frequency when the atom has thermal temperature of $5\ \mu \text{K}$ (blue triangles), $15\ \mu \text{K}$ (red squares), and $50\ \mu \text{K}$ (black circles). The parameters used for the simulation: trap wavelength $\lambda_T=1064$ nm, trap size (beam waist radius) $w_0=2.1\ \mu\text{m}$, trap depth $U_0=1.0$ mK, $\eta=1.68 \times 10^{-4}$ for cesium atom.}
\end{figure}

The fidelity of the $\pi/2$-rotation can be obtained by
\begin{equation}\label{eq14}
F=\mathbf{R}^\mathbf{T} \Theta'^{\mathbf{T}}_{\pi/2} \Theta_{\pi/2} \mathbf{R}
\end{equation}
with $\mathbf{R}$ an arbitrary Bloch vector. Without losing the generality, we use three typical Bloch vectors $(1,0,0)^\mathbf{T}$, $(0,1,0)^\mathbf{T}$, and $(0,0,1)^\mathbf{T}$ to evaluate the Fidelity $F$. Then, we get
\begin{equation}\label{eq15}
F=\frac{1}{3} \left[ \left( \frac{\pi}{2 \theta} \right)^2 \left( 1-\cos{\theta} \right) + \cos{\theta} +\frac{\pi}{\theta} \sin{\theta} \right].
\end{equation}
The overall $\pi/2$-rotation fidelity for the trapped thermal atom is the average over all the vibrational states
\begin{equation}\label{eq16}
\mathbf{F}=\sum_{n_x,n_y,n_z} P_{n_x} P_{n_y} P_{n_z} F.
\end{equation}
Figure~\ref{fig2} shows a simulation of the infidelity, defined by $1-\mathbf{F}$, for a $\pi/2$-rotation versus the Rabi frequency when the atom has different thermal temperatures in a red-detuned ODT. It is obvious that lower temperature and faster rotation give a higher fidelity. If the trapped atom can be cooled to the vibrational ground state, the deteriorating factor on the gate fidelity due to the thermal energy distribution can be totally suppressed. 

\section{Dephasing of the atom}
\subsection{Dephasing due to the thermal distribution on vibrational states}

\begin{figure*}
\includegraphics[width=17 cm]{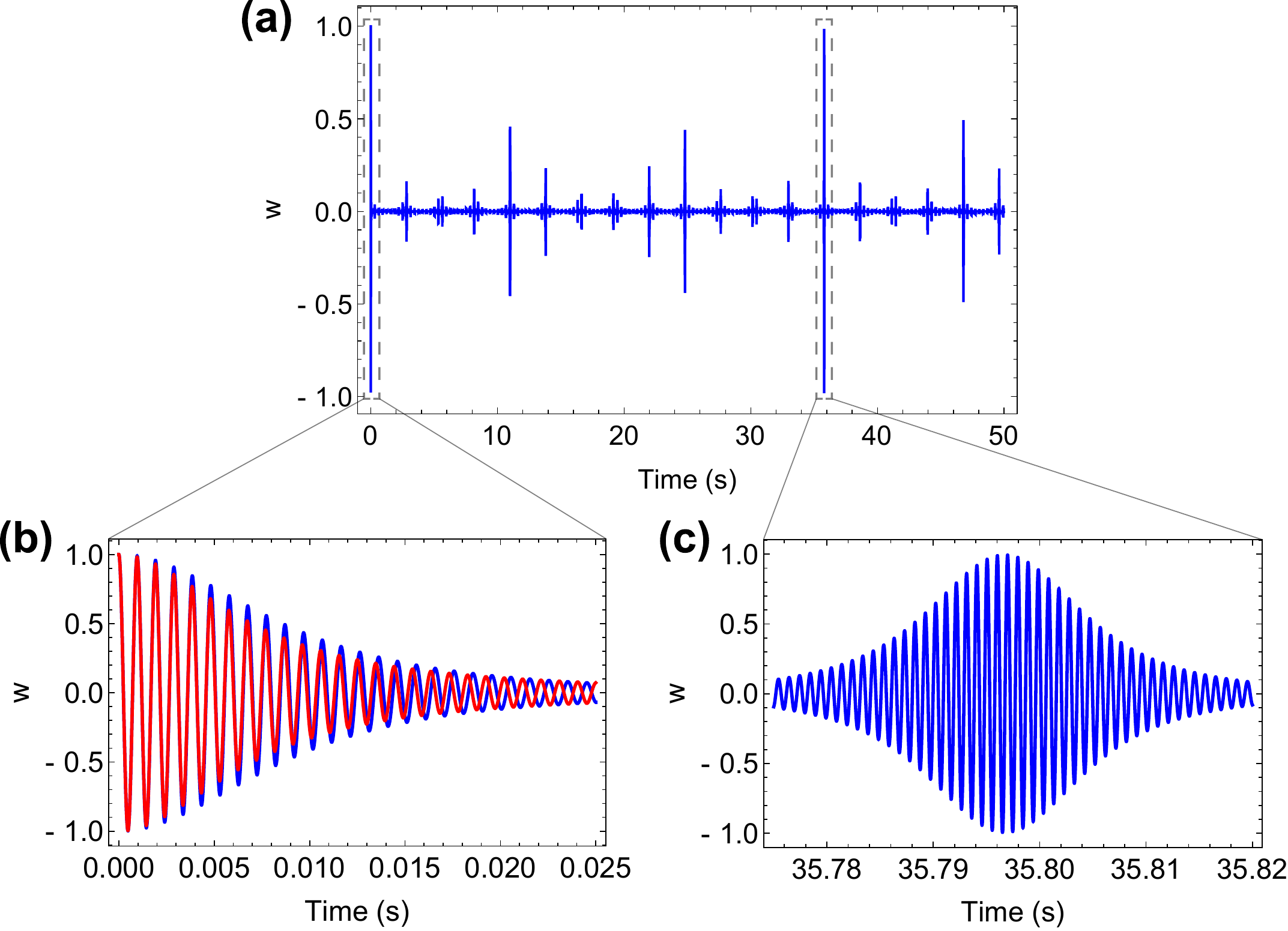}
\caption{\label{fig3} (a)  The simulation of Ramsey fringe of a thermal atom with temperature $15 \ \mu$K by Eq. (\ref{eq19}) with other parameters same as Fig. \ref{fig2}. (b) The zoomed Ramsey fringe within short time delay with blue curve by Eq. (\ref{eq19}) and red curve by Eq. (\ref{eq20a}). (c) The zoomed Ramsey fringe after long time delay where the fringe is fully recovered in the case that no homogeneous dephasing is concerned. }
\end{figure*}

Besides the gate operating error, the thermal distribution on the vibrational states will also induce dephasing due to the variances of energy shift along with different vibrational states. This dephasing mechanism has been classified as inhomogeneous before and since that, in a classical point of view, every single atom possesses different energy and then experience different DLS. Here we show that this mechanism, in a quantum point of view, is actually homogeneous and the mechanism does not induce the dephasing during the state evolution indeed. Usually, the dephasing time can be measured by the Ramsey interference process. If we only consider the dephasing of an atomic coherent superposition state with specific vibrational quantum number $n_x$, $n_y$, and $n_z$, the Ramsey signal will be~\cite{Kuhr2005}
\begin{equation}\label{eq17}
w^\text{Rmsy}_{n}=\cos {\delta_n t}
\end{equation}
with $\delta_n$ the frequency difference between driven field and the atomic transitions and $t$ the free precession time. We have the relation $\delta_n=\delta_0 + \sum_{q=x,y,z} n_q \delta_q$ with $\delta_0$ denotes the frequency difference between driven field and the atomic transition on the ground vibrational state $n_q=0$.  For a trapped atom with thermal energy distribution, the overall Ramsey signal will be the overlap of all the Ramsey signal with the distribution on all the vibrational states:
\begin{equation}\label{eq18}
w^\text{Rmsy}=\sum_{n_x,n_y,n_z} P_{n_x} P_{n_y} P_{n_z} w^\text{Rmsy}_{n}.
\end{equation}
After some algebras, we arrive in
\begin{equation}\label{eq19}
w^\text{Rmsy}=\text{Re}\left[\frac{e^{i \delta_0 t}}{\prod_{q=x,y,z} {\left( \langle n_q \rangle +1 -\langle n_q \rangle e^{i \delta_q t}\right)}} \right]
\end{equation}
with Re means the real part. A typical plot of the Ramsey signal versus time delay is shown in  Fig.~\ref{fig3}. We can see that the amplitude of the Ramsey fringe is not constant. If there is no other homogeneous dephasing factor, the interference will recover at time delay $t=n \times 2 \pi/\delta_\text{GCF} \ (n=1,2,3\cdots)$ with $\delta_\text{GCF}$ the greatest common factor (GCF) of $\delta_x, \ \delta_y, \ \text{and} \ \delta_z$. On short time scale, the amplitude of the fringe drops severely due to the overlap of the interference signal with atom populating on different vibrational states.

Usually, we have the condition $\delta_q \ll \delta_0$ in a red-detuned ODT. Thus, we only consider the time scale $t \ll 1/\delta_q$, Eq. (\ref{eq19}) can be approximated by
\begin{subequations}
\begin{align}
w^\text{Rmsy}&=\frac{\cos{\left[ (\delta_0 +\langle n_x \rangle \delta_x + \langle n_y \rangle \delta_y +\langle n_z \rangle \delta_z) t\right] }}{1 +\left(\langle n_x \rangle \delta_x t\right)^2 + \left(\langle n_y \rangle \delta_y t\right)^2 +\left(\langle n_z \rangle \delta_z t\right)^2 } \label{eq20a}\\
&=\frac{\cos{\left[ (\delta_0 +\langle n_x \rangle \delta_x + \langle n_y \rangle \delta_y +\langle n_z \rangle \delta_z) t\right] }}{1+3\left(\frac{\eta k_B T}{4 \hbar} \right)^2 t^2 }. \label{eq20b}
\end{align}
\end{subequations}
The comparison of Eqs. (\ref{eq19}) and (\ref{eq20b}) under the same parameters are shown in Fig.~\ref{fig3}(b). We can see Eq. (\ref{eq20b}) has included the key features of the Ramsey interference fringe in the short time scale. The dephasing time can be characterized by
\begin{equation}\label{eq21}
T^*_2=2 \sqrt{(e-1)/3} \frac{2 \hbar}{\eta k_B T}=1.51 \frac{2 \hbar}{\eta k_B T},
\end{equation}
in which the dephasing time is defined by the time delay with $1/e$ of the fringe amplitude.

Here we can see that the ``inhomogeneous dephasing mechanism'' due to the thermal motion of the atom in an ODT is actually \textit{homogeneous}, since every single atom experiences the same process. The Ramsey fringe will recover naturally for a long time.

However, the vibrational quantum number will increase due to the heating of the atom by the intensity noise~\cite{Savard1997, Gehm1998}. The heating is random for atom in either of the two fiducial states, and the interference will be eliminated once the original vibrational quantum number is altered stochastically. Due to that the rephasing of a thermal atom populated in different vibrational states usually takes a very long time and the vibrational quantum number would be already changed by the heating process. Thus, the rephasing is hard to be observed in practice.

The observed ``dephasing'' in short time is actually an overlap of the interference signals with a series of different interfere frequencies. Of course, the interference signal can be recovered at any time by using the spin-echo technique to reverse the spin precession direction. By squeezing the atomic distribution on the vibrational state, e.g., cooling the atom into the ground vibrational states, will dramatically suppress the ``dephasing'' since only one frequency would exist in the interference fringes for the ground vibrational state.

\subsection{Dephasing due to the intensity noise of the ODT beam}

Next, we will consider the dephasing due to the intensity noise of trap laser. The intensity noise induces the fluctuation of DFS and causes the dephasing. Assuming there is a small change in trap depth $\delta U_0$ due to the small fluctuation of intensity, we have the variance of the DFS
\begin{equation}\label{eq22}
\delta \Delta^\text{DFS}=-\eta \frac{\delta U_0}{\hbar} \left[ 1 -\sum_{q=x,y,z} \frac{(n_q+\frac{1}{2}) \hbar \omega_q}{2U_0} \right] 
\end{equation}
from Eq. (\ref{eq7}). Usually, the kinetic energy of trapped atom is much lower than the trap depth, so we have $(n_q+\frac{1}{2}) \hbar \omega_q\ll U_0$. Equation (\ref{eq22}) can be reduced to
\begin{equation}\label{eq23}
\delta \Delta^\text{DFS}=-\eta \frac{\delta U_0}{\hbar} \propto \delta I,
\end{equation}
which means that all the trapped atoms take the same dephasing process due to the intensity fluctuations of the trap laser in a red-detuned trap. A Gaussian distribution of fluctuation on trap intensity will also give a Gaussian distribution of DFS. The mean fluctuation of the intensity is $\overline{\delta \Delta^\text{DFS}}=0$ and we have a standard deviation $\sigma^\text{DFS}=\eta \sigma^{U_0} / \hbar $ with $\sigma^{U_0}$ the standard deviation of the trap depth. Then, the theoretical homogeneous dephasing time is
\begin{equation}\label{eq24}
T'_2=\frac{\sqrt{2}}{\sigma^\text{DFS}},
\end{equation}
which is same to the analysis with the atom motion being treated classically~\cite{Kuhr2005}.

\subsection{Dephasing of atom in a blue-detuned trap}
In a blue-detuned trap, the atom is trapped in regions with intensity minima which is surrounded by high potential barriers. A typical trap is the blue-detuned optical lattice (BDOL) by interfering laser beams. The oscillation frequency on one axis for a three dimensional BDOL is $\omega_q=\frac{\sqrt{2} \pi}{d_q} \sqrt{\frac{U_q}{m}}$ with $d_q$ the trap period, $U_q$ the height of trap barrier, and $q=x,\ y,$ or $z$ representing the corresponding axis. If we assume the residual potential at trap site is $U_0$, the overall DFS of the trapped atom still has the form of  Eq. (\ref{eq7}) except that $\delta_q=\left| \frac{\omega_\text{hfs}}{\Delta} \right| \frac{\sqrt{2} \pi}{2 d_q} \sqrt{\frac{U_q}{m}} = \frac{\eta}{2}\omega_{q}$ in the second term.
The ``inhomogeneous dephasing'', roots from the thermal distribution on the vibrational states, and the Ramsey fringe and the dephasing time $T^*_2$ are similar to the scenario in red-detuned trap and with forms as shown in Eqs. (\ref{eq19})-(\ref{eq21}). However, the homogeneous dephasing is very different due to the small residual potential $U_0$ at the trap sites. In a perfectly aligned lattice, we have $U_0=0$, thus the fluctuation of DFS only depends on the height of trap barriers by 
\begin{equation}\label{eq25}
\delta \Delta^\text{DFS}=\eta \sum_{q=x,y,z} \frac{\delta U_q}{\hbar} \frac{(n_q+\frac{1}{2}) \hbar \omega_q}{2U_q},
\end{equation}
which is much smaller than the fluctuations of trap barrier $\delta U_q$. The dephasing time is then much longer than that in a red-detuned trap. This has been proven by our previous experiment ~\cite{Tian2019}.

\section{``Magic'' trapping depth (intensity) }

\begin{figure}
\includegraphics[width=0.75 \columnwidth]{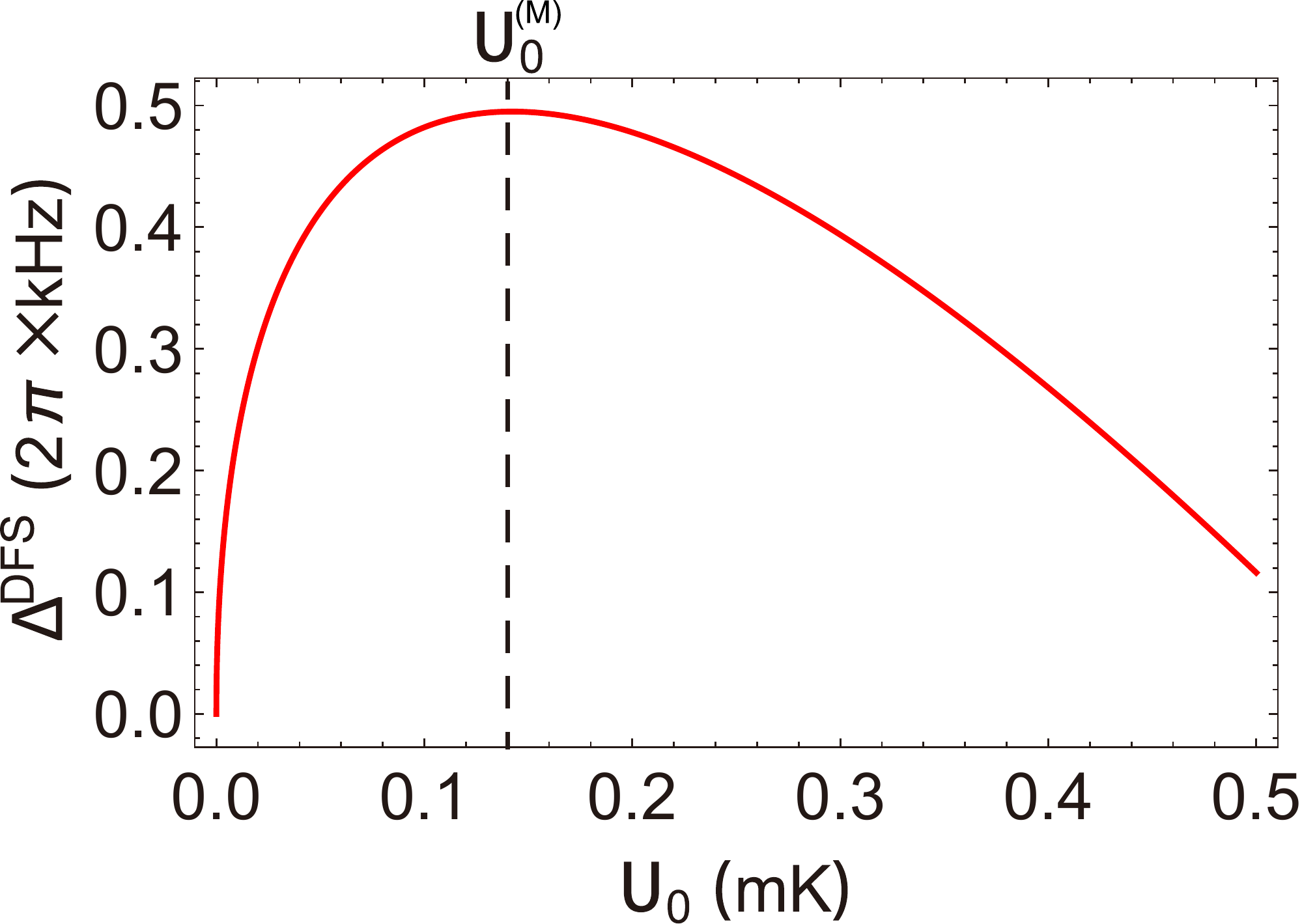}
\caption{\label{fig4} The DFS between $|F_1\rangle$ and $|F_2\rangle$ in a red-detuned trap as a function of trap depth for the atom prepared in vibrational states with $n_x=n_y=300$ and $n_z=2000$. The ``magic'' trap depth $U_0^{(M)}$ can be found at $0.14$ mK.}
\end{figure}  

For a strongly focused red-detuned ODT as discussed in Sec. II, we check Eq.~(\ref{eq7}) again. The first term on the rhs is always negative and proportional to $U_0$, wheras the second term is always positive and proportional to $\sqrt{U_0}$. The overall DFS has a quadric dependence on $\sqrt{U_0}$ (see Fig. \ref{fig4}) and would have a maxima at special value of $U_0^{(M)}$ (the dashed line), where $\partial{\Delta^\text{DFS}}/\partial{U_0}=0$. That also means that the first-order dependence of DFS on $U_0$ will be cancelled at this point. We therefore have 
\begin{equation}\label{eq26}
U_0^{(M)} =A^2 \frac{\hbar^2}{4 m}
\end{equation}
from Eq.~(\ref{eq7}) with $A=\frac{n_x + 1/2}{w_x}+\frac{n_y + 1/2}{w_y}+\frac{n_z+1/2}{\sqrt{2} L_f}$. If the trap depth is set at this point, the residual variance of DLS due to the noise on trap depth (intensity) would be dramatically suppressed. That defined as a ``magic'' point for the trap.

However, when the atom is cooled to the vibrational ground states with $n_x=n_y=n_z=0$ the corresponding ``magic'' trap depth would be $2.3\times 10^{-7}$ mK for the trap parameters used in Fig.~\ref{fig2}. The trap is too shallow to trap the atom. If the vibrational quantum number can be enhanced to $n_x=n_y=300$ and $n_z=2000$, the ``magic'' trap depth can also be boosted to $0.14$ mK. This is deep enough for trapping the atom, but those quantum vibrational states are hard to be precisely prepared.

This dilemma in the red-detuned ODT can be resolved by prepare the fiducial states on different vibrational quantum states. The quantum rotation between the two states can be realized by using the Raman lasers with different wave vectors or shifting the spatial coordinate as being adopted in the Raman sideband cooling~\cite{Forster2009, LiX2012, Kaufman2012, Thompson2013}. We assume that the atom has been prepared into the ground vibrational states. The two fiducial states are chosen as $|F_1\rangle \otimes |n_x=0\rangle$ and $|F_2\rangle \otimes |n_x=1\rangle$, thus the DFS is 
\begin{equation} \label{eq27}
\Delta^\text{DFS} =-\eta \frac{U_0}{\hbar}+\frac{1}{2}( \delta_x + \delta_y + \delta_z) +\omega_x,
\end{equation}
where an extra term $\omega_x$ representing the one phonon energy on $x$ axis appears. From the abovementioned discussion in Sec. I we have $\delta_q \ll \omega_x$ with $q=x,\ y\ \text{or} \ z$ and $\delta_q$ can be omitted, so Eq.~(\ref{eq27}) can be approximated as
\begin{subequations}
\begin{align}
\Delta^\text{DFS} & = -\eta \frac{U_0}{\hbar}+\omega_x \label{eq28a}\\
& = -\eta \frac{U_0}{\hbar} + \frac{2}{w_0} \sqrt{\frac{U_0}{m}}. \label{eq28b}
\end{align}
\end{subequations}
Then, the new ``magic'' trap depth can be obtained as
\begin{equation}\label{eq29}
U_0^{(M)} = \frac{\hbar^2}{m \eta^2 w_0^2}.
\end{equation}
For the red-detuned trap used in Ref. \cite{Kaufman2012} for rubidium-87, an 852-nm ODT with trap size $w_0=0.76$ $\mu$m is used. If $|F_1=1\rangle \otimes |n_x=0\rangle$ and $|F_2=2\rangle \otimes |n_x=1\rangle$ are adopted, the estimated ``magic'' trap depth is $U_0^{(M)}=182$ mK which is unpractical for a real trap. 

Nevertheless, if the quantum state on z axis is considered, the ``magic'' trap depth would be decreased. So, we assume that $|F_1\rangle \otimes |n_z=0\rangle$ and $|F_2\rangle \otimes |n_z=1\rangle$ are adopt, the ``magic'' trap depth is then
\begin{equation}\label{eq30}
U_0^{(M)} = \frac{\hbar^2}{2 m \eta^2 z_R^2}.
\end{equation}
The ``magic'' trap depth $U_0^{(M)}=11.6$ mK can be found for the trap used in Ref. \cite{Kaufman2012}. The trap is still too deep for practically using. However, if the trap size can be increase to $w_0=1.4$ $\mu$m, the ``magic'' trap depth would be decreased to a practical value with $U_0^{(M)}=1.0 $ mK.

The ``magic'' trapping depth can also be found in the blue-detuned ODT. Here a trap potential $U_0$ is assumed in the trapping spot, and the potential barriers on the three directions are $U_q = \alpha_q U_0$ ($q=x,\ y\ \text{or} \ z$), where $\alpha_q$ is the ratio between the barrier height and trap bottom potential with $\alpha_q>1$.  We still consider the BDOL discussed in Sec. IV with the atom being prepared in the vibrational ground state. The DFS between $|F_1\rangle$ and $|F_2\rangle$ is then 
\begin{equation} \label{eq31}
\Delta^\text{DFS} =-\eta \frac{U_0}{\hbar}+\frac{\eta \pi}{2} \sqrt{\frac{U_0}{2 m}} \sum_q \frac{\sqrt{\alpha_q}}{d_q}.
\end{equation}
The ``magic'' trap depth is then
\begin{equation}\label{eq32}
U_0^{(M)} =\frac{\pi^2 \hbar^2}{32 m} B^2 
\end{equation}
with $B=\sum_q \frac{\sqrt{\alpha_q}}{d_q}$. To give an estimation of the number, we use the trap parameters in Ref~\cite{Wang2016}, where a three-dimensional 847.78-nm lattice is used to trap single cesium atoms with trap spacing $d_x=d_y=d_z=5 \ \mu\text{m}$. If we still assume $\alpha_x=\alpha_y=\alpha_z=400$, the corresponding ``magic'' potential at the trap bottom is $U_0^{(M)} =0.16 \ \mu\text{K}$. The height of the trap barrier is then $U_q = \alpha_q U_0 = 65 \ \mu\text{K}$, which is deep enough to trap single atoms.

Here we see that by preparing the atom in special vibrational states, especially in the ground states, a ``magic'' trapping depth could be found. At the ``magic'' point, the DFS is independent from the first-order of trap depth. The residual higher-order terms play minor roles and the dephasing between two fiducial states can be dramatically suppressed. The ``magic'' condition found here is free from the wavelengths and polarization of the trapping beams. Moreover, the magic condition of magnetic field can be applied independently, thus the ``doubly magic'' conditions for both the trap beam and magnetic field are promised.

\section{Conclusion}
In this paper we have analyzed the fidelity of coherent rotating gate and the dephasing between two hyperfine states of a thermal atom trapped in ODT. We treat the atom as a quantum oscillator, instead of a classical oscillator. For a thermal trapped atom, due to the thermal distribution on the vibrational quantum states, the fidelity of a gate is limited by the temperature. The dephasing of the fiducial states is also revised. The dephasing due to the thermal motion of the atom, which has been thought to be inhomogeneous before, is actually homogeneous because that each atom has same distribution on the vibrational states. Moreover, the Ramsey fringe would recover after a longer time due to the rephasing as long as the DLS is stable and the vibrational quantum states are not altered. 

However, in a longer time scale, the atom would be heated up due to the intensity noise of the ODT beam. The stochastic alternation of the vibrational quantum states will destroy the recovery of the fringe. In addition, the variance of DFS due to the intensity fluctuation of trap beam also induces the dephasing and prevents the recovery of the fringe. In a red-detuned ODT, the atom is trapped at the position with local intensity maxima, where the heating of the atom and the fluctuation of DFS are dominated by the local intensity noise. Whereas, in an ideal blue-detuned ODT, the atom is trapped in the spot with zero intensity and the local intensity noise is also eliminated intrinsically. The heating of the atom and the fluctuations of the DFS can be dramatically suppressed. Therefore, the atom in blue-detuned ODT naturally possesses longer coherence time. Significantly, the variance of DFS can also be suppressed by preparing the atom in vibrational ground states and give a long coherence time.

In addition, the two different DFS mechanisms, which are due to the local trap depth and the unequal phonon energy between two fiducial states, give two terms with different dependance of the DFS on the trap depth. Thus, the ``magic'' trap depth, in which the first-order dependance of DLS on trap depth (intensity of ODT beam) is removed, is found. We gave several solutions to get the ``magic'' trap depth for both the red-detuned and blue-detuned ODTs. These conditions have no further requirements on the wavelengths and the polarization ODT beams. The ``magic'' condition of magnetic field can thus be applied independently.

Here in this paper, we focused the DFS between two electronic ground hyperfine states, which are usually used to store a qubit. To proceed the quantum information, the atom is excited to Rydberg states where the long-range interaction can be harnessed to create a two-qubit gate. Due to the positive polarizability of the Rydberg atom to the light field, the Rydberg atom always experience a positive potential. So, it cannot be trapped in a red-detuned ODT. To execute a two-qubit gate, the red-detuned ODT will be switched off for a short time. However, the recent progress on trapping the Rydberg atoms with ponderomotive bottle beam traps (BBTs) \cite{Barredo2020} makes it possible that the two-qubit gate can be executed without switching the trap off. In such a BBT, the ponderomotive energy shift for Rydberg atom with $n\sim 100$ is about $200 \ \mu\text{K}$~\cite{Barredo2020,Zhang2011}, which is much larger than the light shift of the ground state ($\sim 0$) in BBT. The DFS between the Rydberg and ground state is dominated by the first term on the rhs of Eq.~(\ref{eq7}), and the second term related to the vibrational state can be omitted. Whereas, it is possible to compensate the trap induced shift at the trap center by adding an additional constant background field~\cite{Zhang2011}. In this case, the magnitude of the first term on the rhs of Eq.~(\ref{eq7}) can be tuned to be comparable to the second term, and the ``magic'' condition to eliminate first-order dependence of the Rydberg-ground DFS to the noise on the trap intensity is possible. Then, the coherence between Rydberg and ground states can be improved, and the fidelity of the two-qubit gate may also be enhanced. 

This work was supported by the National Key Research and Development Program of China (Grans Nos. 2021YFA1402002 and 2017YFA0304502), the National Natural Science Foundation of China (Grant Nos. U21A6006, U21A20433, 11974223, 11974225, 12104277, and 12104278), and the Fund for Shanxi 1331 Project Key Subjects Construction.


\begin{thebibliography}{42}%
\makeatletter
\providecommand \@ifxundefined [1]{%
 \@ifx{#1\undefined}
}%
\providecommand \@ifnum [1]{%
 \ifnum #1\expandafter \@firstoftwo
 \else \expandafter \@secondoftwo
 \fi
}%
\providecommand \@ifx [1]{%
 \ifx #1\expandafter \@firstoftwo
 \else \expandafter \@secondoftwo
 \fi
}%
\providecommand \natexlab [1]{#1}%
\providecommand \enquote  [1]{``#1''}%
\providecommand \bibnamefont  [1]{#1}%
\providecommand \bibfnamefont [1]{#1}%
\providecommand \citenamefont [1]{#1}%
\providecommand \href@noop [0]{\@secondoftwo}%
\providecommand \href [0]{\begingroup \@sanitize@url \@href}%
\providecommand \@href[1]{\@@startlink{#1}\@@href}%
\providecommand \@@href[1]{\endgroup#1\@@endlink}%
\providecommand \@sanitize@url [0]{\catcode `\\12\catcode `\$12\catcode
  `\&12\catcode `\#12\catcode `\^12\catcode `\_12\catcode `\%12\relax}%
\providecommand \@@startlink[1]{}%
\providecommand \@@endlink[0]{}%
\providecommand \url  [0]{\begingroup\@sanitize@url \@url }%
\providecommand \@url [1]{\endgroup\@href {#1}{\urlprefix }}%
\providecommand \urlprefix  [0]{URL }%
\providecommand \Eprint [0]{\href }%
\providecommand \doibase [0]{http://dx.doi.org/}%
\providecommand \selectlanguage [0]{\@gobble}%
\providecommand \bibinfo  [0]{\@secondoftwo}%
\providecommand \bibfield  [0]{\@secondoftwo}%
\providecommand \translation [1]{[#1]}%
\providecommand \BibitemOpen [0]{}%
\providecommand \bibitemStop [0]{}%
\providecommand \bibitemNoStop [0]{.\EOS\space}%
\providecommand \EOS [0]{\spacefactor3000\relax}%
\providecommand \BibitemShut  [1]{\csname bibitem#1\endcsname}%
\let\auto@bib@innerbib\@empty
%</preamble>
\bibitem [{\citenamefont {Ye}\ \emph {et~al.}(2008)\citenamefont {Ye},
  \citenamefont {Kimble},\ and\ \citenamefont {Katori}}]{Jun2008}%
  \BibitemOpen
  \bibfield  {author} {\bibinfo {author} {\bibfnamefont {J.}~\bibnamefont
  {Ye}}, \bibinfo {author} {\bibfnamefont {H.~J.}\ \bibnamefont {Kimble}}, \
  and\ \bibinfo {author} {\bibfnamefont {H.}~\bibnamefont {Katori}},\
  }\bibfield  {title} {\enquote {\bibinfo {title} {Quantum state engineering
  and precision metrology using state-insensitive light traps},}\ }\href
  {\doibase 10.1126/science.1148259} {\bibfield  {journal} {\bibinfo  {journal}
  {Science}\ }\textbf {\bibinfo {volume} {320}},\ \bibinfo {pages} {1734}
  (\bibinfo {year} {2008})}\BibitemShut {NoStop}%
\bibitem [{\citenamefont {Derevianko}\ and\ \citenamefont
  {Katori}(2011)}]{Derevianko2011}%
  \BibitemOpen
  \bibfield  {author} {\bibinfo {author} {\bibfnamefont {A.}~\bibnamefont
  {Derevianko}}\ and\ \bibinfo {author} {\bibfnamefont {H.}~\bibnamefont
  {Katori}},\ }\bibfield  {title} {\enquote {\bibinfo {title} {Colloquium:
  Physics of optical lattice clocks},}\ }\href {\doibase 10.1103/RevModPhys.83.331} {\bibfield  {journal} {\bibinfo  {journal} {Rev.
  Mod. Phys.}\ }\textbf {\bibinfo {volume} {83}},\ \bibinfo {pages} {331}
  (\bibinfo {year} {2011})}\BibitemShut {NoStop}%
\bibitem [{\citenamefont {Bloch}\ \emph {et~al.}(2008)\citenamefont {Bloch},
  \citenamefont {Dalibard},\ and\ \citenamefont {Zwerger}}]{Bloch2008}%
  \BibitemOpen
  \bibfield  {author} {\bibinfo {author} {\bibfnamefont {I.}~\bibnamefont
  {Bloch}}, \bibinfo {author} {\bibfnamefont {J.}~\bibnamefont {Dalibard}}, \
  and\ \bibinfo {author} {\bibfnamefont {W.}~\bibnamefont {Zwerger}},\
  }\bibfield  {title} {\enquote {\bibinfo {title} {Many-body physics with
  ultracold gases},}\ }\href {\doibase 10.1103/RevModPhys.80.885} {\bibfield
  {journal} {\bibinfo  {journal} {Rev. Mod. Phys.}\ }\textbf {\bibinfo {volume}
  {80}},\ \bibinfo {pages} {885} (\bibinfo {year} {2008})}\BibitemShut
  {NoStop}%
\bibitem [{\citenamefont {Labuhn}\ \emph {et~al.}(2016)\citenamefont {Labuhn},
  \citenamefont {Barredo}, \citenamefont {Ravets}, \citenamefont
  {de~L\'{e}s\'{e}leuc}, \citenamefont {Macr\`{\i}}, \citenamefont {Lahaye},\
  and\ \citenamefont {Browaeys}}]{Labuhn2016}%
  \BibitemOpen
  \bibfield  {author} {\bibinfo {author} {\bibfnamefont {H.}~\bibnamefont
  {Labuhn}}, \bibinfo {author} {\bibfnamefont {D.}~\bibnamefont {Barredo}},
  \bibinfo {author} {\bibfnamefont {S.}~\bibnamefont {Ravets}}, \bibinfo
  {author} {\bibfnamefont {S.}~\bibnamefont {de~L\'{e}s\'{e}leuc}}, \bibinfo
  {author} {\bibfnamefont {T.}~\bibnamefont {Macr\`{\i}}}, \bibinfo {author}
  {\bibfnamefont {T.}~\bibnamefont {Lahaye}}, \ and\ \bibinfo {author}
  {\bibfnamefont {A.}~\bibnamefont {Browaeys}},\ }\bibfield  {title} {\enquote
  {\bibinfo {title} {Tunable two-dimensional arrays of single rydberg atoms for
  realizing quantum ising models},}\ }\href {\doibase 10.1038/nature18274}
  {\bibfield  {journal} {\bibinfo  {journal} {Nature}\ }\textbf {\bibinfo
  {volume} {534}},\ \bibinfo {pages} {667} (\bibinfo {year}
  {2016})}\BibitemShut {NoStop}%
\bibitem [{\citenamefont {Browaeys}\ and\ \citenamefont
  {Lahaye}(2020)}]{Browaeys2020}%
  \BibitemOpen
  \bibfield  {author} {\bibinfo {author} {\bibfnamefont {A.}~\bibnamefont
  {Browaeys}}\ and\ \bibinfo {author} {\bibfnamefont {T.}~\bibnamefont
  {Lahaye}},\ }\bibfield  {title} {\enquote {\bibinfo {title} {Many-body
  physics with individually-controlled rydberg atoms},}\ }\href {\doibase 10.1038/s41567-019-0733-z} {\bibfield  {journal} {\bibinfo  {journal} {Nature
  Physics}\ }\textbf {\bibinfo {volume} {16}},\ \bibinfo {pages} {1} (\bibinfo
  {year} {2020})}\BibitemShut {NoStop}%
\bibitem [{\citenamefont {Saffman}\ \emph {et~al.}(2010)\citenamefont
  {Saffman}, \citenamefont {Walker},\ and\ \citenamefont
  {M\o{}lmer}}]{Saffman2010}%
  \BibitemOpen
  \bibfield  {author} {\bibinfo {author} {\bibfnamefont {M.}~\bibnamefont
  {Saffman}}, \bibinfo {author} {\bibfnamefont {T.~G.}\ \bibnamefont {Walker}},
  \ and\ \bibinfo {author} {\bibfnamefont {K.}~\bibnamefont {M\o{}lmer}},\
  }\bibfield  {title} {\enquote {\bibinfo {title} {Quantum information with
  rydberg atoms},}\ }\href {\doibase 10.1103/RevModPhys.82.2313} {\bibfield
  {journal} {\bibinfo  {journal} {Rev. Mod. Phys.}\ }\textbf {\bibinfo {volume}
  {82}},\ \bibinfo {pages} {2313} (\bibinfo {year} {2010})}\BibitemShut
  {NoStop}%
\bibitem [{\citenamefont {Reiserer}\ and\ \citenamefont
  {Rempe}(2015)}]{Reiserer2015}%
  \BibitemOpen
  \bibfield  {author} {\bibinfo {author} {\bibfnamefont {A.}~\bibnamefont
  {Reiserer}}\ and\ \bibinfo {author} {\bibfnamefont {G.}~\bibnamefont
  {Rempe}},\ }\bibfield  {title} {\enquote {\bibinfo {title} {Cavity-based
  quantum networks with single atoms and optical photons},}\ }\href {\doibase 10.1103/RevModPhys.87.1379} {\bibfield  {journal} {\bibinfo  {journal} {Rev.
  Mod. Phys.}\ }\textbf {\bibinfo {volume} {87}},\ \bibinfo {pages} {1379}
  (\bibinfo {year} {2015})}\BibitemShut {NoStop}%
\bibitem [{\citenamefont {Lundblad}\ \emph {et~al.}(2010)\citenamefont
  {Lundblad}, \citenamefont {Schlosser},\ and\ \citenamefont
  {Porto}}]{Lundblad2010}%
  \BibitemOpen
  \bibfield  {author} {\bibinfo {author} {\bibfnamefont {N.}~\bibnamefont
  {Lundblad}}, \bibinfo {author} {\bibfnamefont {M.}~\bibnamefont {Schlosser}},
  \ and\ \bibinfo {author} {\bibfnamefont {J.~V.}\ \bibnamefont {Porto}},\
  }\bibfield  {title} {\enquote {\bibinfo {title} {Experimental observation of
  magic-wavelength behavior of $^{87}\mathrm{Rb}$ atoms in an optical
  lattice},}\ }\href {\doibase 10.1103/PhysRevA.81.031611} {\bibfield
  {journal} {\bibinfo  {journal} {Phys. Rev. A}\ }\textbf {\bibinfo {volume}
  {81}},\ \bibinfo {pages} {031611} (\bibinfo {year} {2010})}\BibitemShut
  {NoStop}%
\bibitem [{\citenamefont {Dudin}\ \emph {et~al.}(2010)\citenamefont {Dudin},
  \citenamefont {Zhao}, \citenamefont {Kennedy},\ and\ \citenamefont
  {Kuzmich}}]{Dudin2010}%
  \BibitemOpen
  \bibfield  {author} {\bibinfo {author} {\bibfnamefont {Y.~O.}\ \bibnamefont
  {Dudin}}, \bibinfo {author} {\bibfnamefont {R.}~\bibnamefont {Zhao}},
  \bibinfo {author} {\bibfnamefont {T.~A.~B.}\ \bibnamefont {Kennedy}}, \ and\
  \bibinfo {author} {\bibfnamefont {A.}~\bibnamefont {Kuzmich}},\ }\bibfield
  {title} {\enquote {\bibinfo {title} {Light storage in a magnetically dressed
  optical lattice},}\ }\href {\doibase 10.1103/PhysRevA.81.041805} {\bibfield
  {journal} {\bibinfo  {journal} {Phys. Rev. A}\ }\textbf {\bibinfo {volume}
  {81}},\ \bibinfo {pages} {041805} (\bibinfo {year} {2010})}\BibitemShut
  {NoStop}%
\bibitem [{\citenamefont {Kim}\ \emph {et~al.}(2013)\citenamefont {Kim},
  \citenamefont {Han},\ and\ \citenamefont {Cho}}]{Kim2013}%
  \BibitemOpen
  \bibfield  {author} {\bibinfo {author} {\bibfnamefont {H.}~\bibnamefont
  {Kim}}, \bibinfo {author} {\bibfnamefont {H.~S.}\ \bibnamefont {Han}}, \ and\
  \bibinfo {author} {\bibfnamefont {D.}~\bibnamefont {Cho}},\ }\bibfield
  {title} {\enquote {\bibinfo {title} {Magic polarization for optical trapping
  of atoms without stark-induced dephasing},}\ }\href {\doibase 10.1103/PhysRevLett.111.243004} {\bibfield  {journal} {\bibinfo  {journal}
  {Phys. Rev. Lett.}\ }\textbf {\bibinfo {volume} {111}},\ \bibinfo {pages}
  {243004} (\bibinfo {year} {2013})}\BibitemShut {NoStop}%
\bibitem [{\citenamefont {Kazakov}\ and\ \citenamefont
  {Schumm}(2015)}]{Kazakov2015}%
  \BibitemOpen
  \bibfield  {author} {\bibinfo {author} {\bibfnamefont {G.~A.}\ \bibnamefont
  {Kazakov}}\ and\ \bibinfo {author} {\bibfnamefont {T.}~\bibnamefont
  {Schumm}},\ }\bibfield  {title} {\enquote {\bibinfo {title} {Magic
  radio-frequency dressing for trapped atomic microwave clocks},}\ }\href
  {\doibase 10.1103/PhysRevA.91.023404} {\bibfield  {journal} {\bibinfo
  {journal} {Phys. Rev. A}\ }\textbf {\bibinfo {volume} {91}},\ \bibinfo
  {pages} {023404} (\bibinfo {year} {2015})}\BibitemShut {NoStop}%
\bibitem [{\citenamefont {Yang}\ \emph {et~al.}(2016)\citenamefont {Yang},
  \citenamefont {He}, \citenamefont {Guo}, \citenamefont {Xu}, \citenamefont
  {Wang}, \citenamefont {Sheng}, \citenamefont {Liu}, \citenamefont {Wang},
  \citenamefont {Derevianko},\ and\ \citenamefont {Zhan}}]{Yang2016}%
  \BibitemOpen
  \bibfield  {author} {\bibinfo {author} {\bibfnamefont {J.}~\bibnamefont
  {Yang}}, \bibinfo {author} {\bibfnamefont {X.}~\bibnamefont {He}}, \bibinfo
  {author} {\bibfnamefont {R.}~\bibnamefont {Guo}}, \bibinfo {author}
  {\bibfnamefont {P.}~\bibnamefont {Xu}}, \bibinfo {author} {\bibfnamefont
  {K.}~\bibnamefont {Wang}}, \bibinfo {author} {\bibfnamefont {C.}~\bibnamefont
  {Sheng}}, \bibinfo {author} {\bibfnamefont {M.}~\bibnamefont {Liu}}, \bibinfo
  {author} {\bibfnamefont {J.}~\bibnamefont {Wang}}, \bibinfo {author}
  {\bibfnamefont {A.}~\bibnamefont {Derevianko}}, \ and\ \bibinfo {author}
  {\bibfnamefont {M.}~\bibnamefont {Zhan}},\ }\bibfield  {title} {\enquote
  {\bibinfo {title} {Coherence preservation of a single neutral atom qubit
  transferred between magic-intensity optical traps},}\ }\href {\doibase 10.1103/PhysRevLett.117.123201} {\bibfield  {journal} {\bibinfo  {journal}
  {Phys. Rev. Lett.}\ }\textbf {\bibinfo {volume} {117}},\ \bibinfo {pages}
  {123201} (\bibinfo {year} {2016})}\BibitemShut {NoStop}%
\bibitem [{\citenamefont {Flambaum}\ \emph {et~al.}(2008)\citenamefont
  {Flambaum}, \citenamefont {Dzuba},\ and\ \citenamefont
  {Derevianko}}]{Flambaum2008}%
  \BibitemOpen
  \bibfield  {author} {\bibinfo {author} {\bibfnamefont {V.~V.}\ \bibnamefont
  {Flambaum}}, \bibinfo {author} {\bibfnamefont {V.~A.}\ \bibnamefont {Dzuba}},
  \ and\ \bibinfo {author} {\bibfnamefont {A.}~\bibnamefont {Derevianko}},\
  }\bibfield  {title} {\enquote {\bibinfo {title} {Magic frequencies for cesium
  primary-frequency standard},}\ }\href {\doibase 10.1103/PhysRevLett.101.220801} {\bibfield  {journal} {\bibinfo  {journal}
  {Phys. Rev. Lett.}\ }\textbf {\bibinfo {volume} {101}},\ \bibinfo {pages}
  {220801} (\bibinfo {year} {2008})}\BibitemShut {NoStop}%
\bibitem [{\citenamefont {Derevianko}(2010)}]{Derevianko2010}%
  \BibitemOpen
  \bibfield  {author} {\bibinfo {author} {\bibfnamefont {A.}~\bibnamefont
  {Derevianko}},\ }\bibfield  {title} {\enquote {\bibinfo {title} {``doubly
  magic'' conditions in magic-wavelength trapping of ultracold alkali-metal
  atoms},}\ }\href {\doibase 10.1103/PhysRevLett.105.033002} {\bibfield
  {journal} {\bibinfo  {journal} {Phys. Rev. Lett.}\ }\textbf {\bibinfo
  {volume} {105}},\ \bibinfo {pages} {033002} (\bibinfo {year}
  {2010})}\BibitemShut {NoStop}%
\bibitem [{\citenamefont {Chicireanu}\ \emph {et~al.}(2011)\citenamefont
  {Chicireanu}, \citenamefont {Nelson}, \citenamefont {Olmschenk},
  \citenamefont {Lundblad}, \citenamefont {Derevianko},\ and\ \citenamefont
  {Porto}}]{Chicireanu2011}%
  \BibitemOpen
  \bibfield  {author} {\bibinfo {author} {\bibfnamefont {R.}~\bibnamefont
  {Chicireanu}}, \bibinfo {author} {\bibfnamefont {K.~D.}\ \bibnamefont
  {Nelson}}, \bibinfo {author} {\bibfnamefont {S.}~\bibnamefont {Olmschenk}},
  \bibinfo {author} {\bibfnamefont {N.}~\bibnamefont {Lundblad}}, \bibinfo
  {author} {\bibfnamefont {A.}~\bibnamefont {Derevianko}}, \ and\ \bibinfo
  {author} {\bibfnamefont {J.~V.}\ \bibnamefont {Porto}},\ }\bibfield  {title}
  {\enquote {\bibinfo {title} {Differential light-shift cancellation in a
  magnetic-field-insensitive transition of $^{87}\mathrm{Rb}$},}\ }\href
  {\doibase 10.1103/PhysRevLett.106.063002} {\bibfield  {journal} {\bibinfo
  {journal} {Phys. Rev. Lett.}\ }\textbf {\bibinfo {volume} {106}},\ \bibinfo
  {pages} {063002} (\bibinfo {year} {2011})}\BibitemShut {NoStop}%
\bibitem [{\citenamefont {Carr}\ and\ \citenamefont
  {Saffman}(2016)}]{Carr2016}%
  \BibitemOpen
  \bibfield  {author} {\bibinfo {author} {\bibfnamefont {A.~W.}\ \bibnamefont
  {Carr}}\ and\ \bibinfo {author} {\bibfnamefont {M.}~\bibnamefont {Saffman}},\
  }\bibfield  {title} {\enquote {\bibinfo {title} {Doubly magic optical
  trapping for cs atom hyperfine clock transitions},}\ }\href {\doibase 10.1103/PhysRevLett.117.150801} {\bibfield  {journal} {\bibinfo  {journal}
  {Phys. Rev. Lett.}\ }\textbf {\bibinfo {volume} {117}},\ \bibinfo {pages}
  {150801} (\bibinfo {year} {2016})}\BibitemShut {NoStop}%
\bibitem [{\citenamefont {Li}\ \emph {et~al.}(2019)\citenamefont {Li},
  \citenamefont {Tian}, \citenamefont {Wu}, \citenamefont {Li}, \citenamefont
  {Li}, \citenamefont {Liu}, \citenamefont {Zhang},\ and\ \citenamefont
  {Zhang}}]{Li2019}%
  \BibitemOpen
  \bibfield  {author} {\bibinfo {author} {\bibfnamefont {G.}~\bibnamefont
  {Li}}, \bibinfo {author} {\bibfnamefont {Y.}~\bibnamefont {Tian}}, \bibinfo
  {author} {\bibfnamefont {W.}~\bibnamefont {Wu}}, \bibinfo {author}
  {\bibfnamefont {S.}~\bibnamefont {Li}}, \bibinfo {author} {\bibfnamefont
  {X.}~\bibnamefont {Li}}, \bibinfo {author} {\bibfnamefont {Y.}~\bibnamefont
  {Liu}}, \bibinfo {author} {\bibfnamefont {P.}~\bibnamefont {Zhang}}, \ and\
  \bibinfo {author} {\bibfnamefont {T.}~\bibnamefont {Zhang}},\ }\bibfield
  {title} {\enquote {\bibinfo {title} {Triply magic conditions for microwave
  transition of optically trapped alkali-metal atoms},}\ }\href {\doibase 10.1103/PhysRevLett.123.253602} {\bibfield  {journal} {\bibinfo  {journal}
  {Phys. Rev. Lett.}\ }\textbf {\bibinfo {volume} {123}},\ \bibinfo {pages}
  {253602} (\bibinfo {year} {2019})}\BibitemShut {NoStop}%
\bibitem [{\citenamefont {Xia}\ \emph {et~al.}(2015)\citenamefont {Xia},
  \citenamefont {Lichtman}, \citenamefont {Maller}, \citenamefont {Carr},
  \citenamefont {Piotrowicz}, \citenamefont {Isenhower},\ and\ \citenamefont
  {Saffman}}]{Xia2015}%
  \BibitemOpen
  \bibfield  {author} {\bibinfo {author} {\bibfnamefont {T.}~\bibnamefont
  {Xia}}, \bibinfo {author} {\bibfnamefont {M.}~\bibnamefont {Lichtman}},
  \bibinfo {author} {\bibfnamefont {K.}~\bibnamefont {Maller}}, \bibinfo
  {author} {\bibfnamefont {A.~W.}\ \bibnamefont {Carr}}, \bibinfo {author}
  {\bibfnamefont {M.~J.}\ \bibnamefont {Piotrowicz}}, \bibinfo {author}
  {\bibfnamefont {L.}~\bibnamefont {Isenhower}}, \ and\ \bibinfo {author}
  {\bibfnamefont {M.}~\bibnamefont {Saffman}},\ }\bibfield  {title} {\enquote
  {\bibinfo {title} {Randomized benchmarking of single-qubit gates in a 2d
  array of neutral-atom qubits},}\ }\href {\doibase 10.1103/PhysRevLett.114.100503} {\bibfield  {journal} {\bibinfo  {journal}
  {Phys. Rev. Lett.}\ }\textbf {\bibinfo {volume} {114}},\ \bibinfo {pages}
  {100503} (\bibinfo {year} {2015})}\BibitemShut {NoStop}%
\bibitem [{\citenamefont {Sheng}\ \emph {et~al.}(2018)\citenamefont {Sheng},
  \citenamefont {He}, \citenamefont {Xu}, \citenamefont {Guo}, \citenamefont
  {Wang}, \citenamefont {Xiong}, \citenamefont {Liu}, \citenamefont {Wang},\
  and\ \citenamefont {Zhan}}]{Sheng2018}%
  \BibitemOpen
  \bibfield  {author} {\bibinfo {author} {\bibfnamefont {C.}~\bibnamefont
  {Sheng}}, \bibinfo {author} {\bibfnamefont {X.}~\bibnamefont {He}}, \bibinfo
  {author} {\bibfnamefont {P.}~\bibnamefont {Xu}}, \bibinfo {author}
  {\bibfnamefont {R.}~\bibnamefont {Guo}}, \bibinfo {author} {\bibfnamefont
  {K.}~\bibnamefont {Wang}}, \bibinfo {author} {\bibfnamefont {Z.}~\bibnamefont
  {Xiong}}, \bibinfo {author} {\bibfnamefont {M.}~\bibnamefont {Liu}}, \bibinfo
  {author} {\bibfnamefont {J.}~\bibnamefont {Wang}}, \ and\ \bibinfo {author}
  {\bibfnamefont {M.}~\bibnamefont {Zhan}},\ }\bibfield  {title} {\enquote
  {\bibinfo {title} {High-fidelity single-qubit gates on neutral atoms in a
  two-dimensional magic-intensity optical dipole trap array},}\ }\href
  {\doibase 10.1103/PhysRevLett.121.240501} {\bibfield  {journal} {\bibinfo
  {journal} {Phys. Rev. Lett.}\ }\textbf {\bibinfo {volume} {121}},\ \bibinfo
  {pages} {240501} (\bibinfo {year} {2018})}\BibitemShut {NoStop}%
\bibitem [{\citenamefont {Guo}\ \emph {et~al.}(2020)\citenamefont {Guo},
  \citenamefont {He}, \citenamefont {Sheng}, \citenamefont {Yang},
  \citenamefont {Xu}, \citenamefont {Wang}, \citenamefont {Zhong},
  \citenamefont {Liu}, \citenamefont {Wang},\ and\ \citenamefont
  {Zhan}}]{Guo2020}%
  \BibitemOpen
  \bibfield  {author} {\bibinfo {author} {\bibfnamefont {R.}~\bibnamefont
  {Guo}}, \bibinfo {author} {\bibfnamefont {X.}~\bibnamefont {He}}, \bibinfo
  {author} {\bibfnamefont {C.}~\bibnamefont {Sheng}}, \bibinfo {author}
  {\bibfnamefont {J.}~\bibnamefont {Yang}}, \bibinfo {author} {\bibfnamefont
  {P.}~\bibnamefont {Xu}}, \bibinfo {author} {\bibfnamefont {K.}~\bibnamefont
  {Wang}}, \bibinfo {author} {\bibfnamefont {J.}~\bibnamefont {Zhong}},
  \bibinfo {author} {\bibfnamefont {M.}~\bibnamefont {Liu}}, \bibinfo {author}
  {\bibfnamefont {J.}~\bibnamefont {Wang}}, \ and\ \bibinfo {author}
  {\bibfnamefont {M.}~\bibnamefont {Zhan}},\ }\bibfield  {title} {\enquote
  {\bibinfo {title} {Balanced coherence times of atomic qubits of different
  species in a dual $3\ifmmode\times\else\texttimes\fi{}3$ magic-intensity
  optical dipole trap array},}\ }\href {\doibase 10.1103/PhysRevLett.124.153201} {\bibfield  {journal} {\bibinfo  {journal}
  {Phys. Rev. Lett.}\ }\textbf {\bibinfo {volume} {124}},\ \bibinfo {pages}
  {153201} (\bibinfo {year} {2020})}\BibitemShut {NoStop}%
\bibitem [{\citenamefont {{Wu}}\ \emph {et~al.}(2019)\citenamefont {{Wu}},
  \citenamefont {{Kumar}}, \citenamefont {{Giraldo}},\ and\ \citenamefont
  {{Weiss}}}]{Wu2019}%
  \BibitemOpen
  \bibfield  {author} {\bibinfo {author} {\bibfnamefont {T.-Y.}\ \bibnamefont
  {{Wu}}}, \bibinfo {author} {\bibfnamefont {A.}~\bibnamefont {{Kumar}}},
  \bibinfo {author} {\bibfnamefont {F.}~\bibnamefont {{Giraldo}}}, \ and\
  \bibinfo {author} {\bibfnamefont {D.~S.}\ \bibnamefont {{Weiss}}},\
  }\bibfield  {title} {\enquote {\bibinfo {title} {Stern-gerlach detection of
  neutral-atom qubits in a state-dependent optical lattice},}\ }\href {\doibase 10.1038/s41567-019-0478-8} {\bibfield  {journal} {\bibinfo  {journal} {Nat.
  Phys.}\ }\textbf {\bibinfo {volume} {15}},\ \bibinfo {pages} {538} (\bibinfo
  {year} {2019})}\BibitemShut {NoStop}%
\bibitem [{\citenamefont {Li}\ \emph {et~al.}(2020)\citenamefont {Li},
  \citenamefont {Wang}, \citenamefont {Li}, \citenamefont {Tian}, \citenamefont
  {Li}, \citenamefont {Zhang},\ and\ \citenamefont {Zhang}}]{Li2020}%
  \BibitemOpen
  \bibfield  {author} {\bibinfo {author} {\bibfnamefont {X.-Y.}\ \bibnamefont
  {Li}}, \bibinfo {author} {\bibfnamefont {Z.-H.}\ \bibnamefont {Wang}},
  \bibinfo {author} {\bibfnamefont {S.-K.}\ \bibnamefont {Li}}, \bibinfo
  {author} {\bibfnamefont {Y.-L.}\ \bibnamefont {Tian}}, \bibinfo {author}
  {\bibfnamefont {G.}~\bibnamefont {Li}}, \bibinfo {author} {\bibfnamefont
  {P.-F.}\ \bibnamefont {Zhang}}, \ and\ \bibinfo {author} {\bibfnamefont
  {T.-C.}\ \bibnamefont {Zhang}},\ }\bibfield  {title} {\enquote {\bibinfo
  {title} {Measurement of magnetically insensitive state coherent time in blue
  dipole trap},}\ }\href {\doibase 10.7498/aps.69.20192001} {\bibfield
  {journal} {\bibinfo  {journal} {Acta Phys. Sin.}\ }\textbf {\bibinfo {volume}
  {69}},\ \bibinfo {pages} {6} (\bibinfo {year} {2020})}\BibitemShut {NoStop}%
\bibitem [{\citenamefont {Kuhr}\ \emph {et~al.}(2005)\citenamefont {Kuhr},
  \citenamefont {Alt}, \citenamefont {Schrader}, \citenamefont {Dotsenko},
  \citenamefont {Miroshnychenko}, \citenamefont {Rauschenbeutel},\ and\
  \citenamefont {Meschede}}]{Kuhr2005}%
  \BibitemOpen
  \bibfield  {author} {\bibinfo {author} {\bibfnamefont {S.}~\bibnamefont
  {Kuhr}}, \bibinfo {author} {\bibfnamefont {W.}~\bibnamefont {Alt}}, \bibinfo
  {author} {\bibfnamefont {D.}~\bibnamefont {Schrader}}, \bibinfo {author}
  {\bibfnamefont {I.}~\bibnamefont {Dotsenko}}, \bibinfo {author}
  {\bibfnamefont {Y.}~\bibnamefont {Miroshnychenko}}, \bibinfo {author}
  {\bibfnamefont {A.}~\bibnamefont {Rauschenbeutel}}, \ and\ \bibinfo {author}
  {\bibfnamefont {D.}~\bibnamefont {Meschede}},\ }\bibfield  {title} {\enquote
  {\bibinfo {title} {Analysis of dephasing mechanisms in a standing-wave dipole
  trap},}\ }\href {\doibase 10.1103/PhysRevA.72.023406} {\bibfield  {journal}
  {\bibinfo  {journal} {Phys. Rev. A}\ }\textbf {\bibinfo {volume} {72}},\
  \bibinfo {pages} {023406} (\bibinfo {year} {2005})}\BibitemShut {NoStop}%
\bibitem [{\citenamefont {Rosenfeld}\ \emph {et~al.}(2011)\citenamefont
  {Rosenfeld}, \citenamefont {Volz}, \citenamefont {Weber},\ and\ \citenamefont
  {Weinfurter}}]{Rosenfeld2011}%
  \BibitemOpen
  \bibfield  {author} {\bibinfo {author} {\bibfnamefont {W.}~\bibnamefont
  {Rosenfeld}}, \bibinfo {author} {\bibfnamefont {J.}~\bibnamefont {Volz}},
  \bibinfo {author} {\bibfnamefont {M.}~\bibnamefont {Weber}}, \ and\ \bibinfo
  {author} {\bibfnamefont {H.}~\bibnamefont {Weinfurter}},\ }\bibfield  {title}
  {\enquote {\bibinfo {title} {Coherence of a qubit stored in zeeman levels of
  a single optically trapped atom},}\ }\href {\doibase 10.1103/PhysRevA.84.022343} {\bibfield  {journal} {\bibinfo  {journal} {Phys.
  Rev. A}\ }\textbf {\bibinfo {volume} {84}},\ \bibinfo {pages} {022343}
  (\bibinfo {year} {2011})}\BibitemShut {NoStop}%
\bibitem [{\citenamefont {Gerasimov}\ \emph {et~al.}(2021)\citenamefont
  {Gerasimov}, \citenamefont {Yusupov}, \citenamefont {Bobrov}, \citenamefont
  {Shchepanovich}, \citenamefont {Kovlakov}, \citenamefont {Straupe},
  \citenamefont {Kulik},\ and\ \citenamefont {Kupriyanov}}]{Gerasimov2021}%
  \BibitemOpen
  \bibfield  {author} {\bibinfo {author} {\bibfnamefont {L.~V.}\ \bibnamefont
  {Gerasimov}}, \bibinfo {author} {\bibfnamefont {R.~R.}\ \bibnamefont
  {Yusupov}}, \bibinfo {author} {\bibfnamefont {I.~B.}\ \bibnamefont {Bobrov}},
  \bibinfo {author} {\bibfnamefont {D.}~\bibnamefont {Shchepanovich}}, \bibinfo
  {author} {\bibfnamefont {E.~V.}\ \bibnamefont {Kovlakov}}, \bibinfo {author}
  {\bibfnamefont {S.~S.}\ \bibnamefont {Straupe}}, \bibinfo {author}
  {\bibfnamefont {S.~P.}\ \bibnamefont {Kulik}}, \ and\ \bibinfo {author}
  {\bibfnamefont {D.~V.}\ \bibnamefont {Kupriyanov}},\ }\bibfield  {title}
  {\enquote {\bibinfo {title} {Dynamics of a spin qubit in an optical dipole
  trap},}\ }\href {\doibase 10.1103/PhysRevA.103.062426} {\bibfield  {journal}
  {\bibinfo  {journal} {Phys. Rev. A}\ }\textbf {\bibinfo {volume} {103}},\
  \bibinfo {pages} {062426} (\bibinfo {year} {2021})}\BibitemShut {NoStop}%
\bibitem [{\citenamefont {Hahn}(1950)}]{Hahn1950}%
  \BibitemOpen
  \bibfield  {author} {\bibinfo {author} {\bibfnamefont {E.~L.}\ \bibnamefont
  {Hahn}},\ }\bibfield  {title} {\enquote {\bibinfo {title} {Spin echoes},}\
  }\href {\doibase 10.1103/PhysRev.80.580} {\bibfield  {journal} {\bibinfo
  {journal} {Phys. Rev.}\ }\textbf {\bibinfo {volume} {80}},\ \bibinfo {pages}
  {580} (\bibinfo {year} {1950})}\BibitemShut {NoStop}%
\bibitem [{\citenamefont {Andersen}\ \emph {et~al.}(2003)\citenamefont
  {Andersen}, \citenamefont {Kaplan},\ and\ \citenamefont
  {Davidson}}]{Andersen2003}%
  \BibitemOpen
  \bibfield  {author} {\bibinfo {author} {\bibfnamefont {M.~F.}\ \bibnamefont
  {Andersen}}, \bibinfo {author} {\bibfnamefont {A.}~\bibnamefont {Kaplan}}, \
  and\ \bibinfo {author} {\bibfnamefont {N.}~\bibnamefont {Davidson}},\
  }\bibfield  {title} {\enquote {\bibinfo {title} {Echo spectroscopy and
  quantum stability of trapped atoms},}\ }\href {\doibase 10.1103/PhysRevLett.90.023001} {\bibfield  {journal} {\bibinfo  {journal}
  {Phys. Rev. Lett.}\ }\textbf {\bibinfo {volume} {90}},\ \bibinfo {pages}
  {023001} (\bibinfo {year} {2003})}\BibitemShut {NoStop}%
\bibitem [{\citenamefont {Kuhr}\ \emph {et~al.}(2003)\citenamefont {Kuhr},
  \citenamefont {Alt}, \citenamefont {Schrader}, \citenamefont {Dotsenko},
  \citenamefont {Miroshnychenko}, \citenamefont {Rosenfeld}, \citenamefont
  {Khudaverdyan}, \citenamefont {Gomer}, \citenamefont {Rauschenbeutel},\ and\
  \citenamefont {Meschede}}]{Kuhr2003}%
  \BibitemOpen
  \bibfield  {author} {\bibinfo {author} {\bibfnamefont {S.}~\bibnamefont
  {Kuhr}}, \bibinfo {author} {\bibfnamefont {W.}~\bibnamefont {Alt}}, \bibinfo
  {author} {\bibfnamefont {D.}~\bibnamefont {Schrader}}, \bibinfo {author}
  {\bibfnamefont {I.}~\bibnamefont {Dotsenko}}, \bibinfo {author}
  {\bibfnamefont {Y.}~\bibnamefont {Miroshnychenko}}, \bibinfo {author}
  {\bibfnamefont {W.}~\bibnamefont {Rosenfeld}}, \bibinfo {author}
  {\bibfnamefont {M.}~\bibnamefont {Khudaverdyan}}, \bibinfo {author}
  {\bibfnamefont {V.}~\bibnamefont {Gomer}}, \bibinfo {author} {\bibfnamefont
  {A.}~\bibnamefont {Rauschenbeutel}}, \ and\ \bibinfo {author} {\bibfnamefont
  {D.}~\bibnamefont {Meschede}},\ }\bibfield  {title} {\enquote {\bibinfo
  {title} {Coherence properties and quantum state transportation in an optical
  conveyor belt},}\ }\href {\doibase 10.1103/PhysRevLett.91.213002} {\bibfield
  {journal} {\bibinfo  {journal} {Phys. Rev. Lett.}\ }\textbf {\bibinfo
  {volume} {91}},\ \bibinfo {pages} {213002} (\bibinfo {year}
  {2003})}\BibitemShut {NoStop}%
\bibitem [{\citenamefont {Grimm}\ \emph {et~al.}(2000)\citenamefont {Grimm},
  \citenamefont {Weidemüller},\ and\ \citenamefont {Ovchinnikov}}]{Grimm2000}%
  \BibitemOpen
  \bibfield  {author} {\bibinfo {author} {\bibfnamefont {R.}~\bibnamefont
  {Grimm}}, \bibinfo {author} {\bibfnamefont {M.}~\bibnamefont {Weidemüller}},
  \ and\ \bibinfo {author} {\bibfnamefont {Y.}~\bibnamefont {Ovchinnikov}},\
  }\bibfield  {title} {\enquote {\bibinfo {title} {Optical dipole traps for
  neutral atoms},}\ }\href {\doibase 10.1016/S1049-250X(08)60186-X} {\bibfield
  {journal} {\bibinfo  {journal} {Adv. At. Mol. Opt. Phys.}\ }\textbf {\bibinfo
  {volume} {42}},\ \bibinfo {pages} {95} (\bibinfo {year} {2000})}\BibitemShut
  {NoStop}%
\bibitem [{\citenamefont {Schlosser}\ \emph {et~al.}(2002)\citenamefont
  {Schlosser}, \citenamefont {Reymond},\ and\ \citenamefont
  {Grangier}}]{Schlosser2002}%
  \BibitemOpen
  \bibfield  {author} {\bibinfo {author} {\bibfnamefont {N.}~\bibnamefont
  {Schlosser}}, \bibinfo {author} {\bibfnamefont {G.}~\bibnamefont {Reymond}},
  \ and\ \bibinfo {author} {\bibfnamefont {P.}~\bibnamefont {Grangier}},\
  }\bibfield  {title} {\enquote {\bibinfo {title} {Collisional blockade in
  microscopic optical dipole traps},}\ }\href {\doibase 10.1103/PhysRevLett.89.023005} {\bibfield  {journal} {\bibinfo  {journal}
  {Phys. Rev. Lett.}\ }\textbf {\bibinfo {volume} {89}},\ \bibinfo {pages}
  {023005} (\bibinfo {year} {2002})}\BibitemShut {NoStop}%
\bibitem [{\citenamefont {Saffman}\ and\ \citenamefont
  {Walker}(2005)}]{Saffman2005}%
  \BibitemOpen
  \bibfield  {author} {\bibinfo {author} {\bibfnamefont {M.}~\bibnamefont
  {Saffman}}\ and\ \bibinfo {author} {\bibfnamefont {T.~G.}\ \bibnamefont
  {Walker}},\ }\bibfield  {title} {\enquote {\bibinfo {title} {Analysis of a
  quantum logic device based on dipole-dipole interactions of optically trapped
  rydberg atoms},}\ }\href {\doibase 10.1103/PhysRevA.72.022347} {\bibfield
  {journal} {\bibinfo  {journal} {Phys. Rev. A}\ }\textbf {\bibinfo {volume}
  {72}},\ \bibinfo {pages} {022347} (\bibinfo {year} {2005})}\BibitemShut
  {NoStop}%
\bibitem [{\citenamefont {Foot}(2004)}]{Foot2004}%
  \BibitemOpen
  \bibfield  {author} {\bibinfo {author} {\bibfnamefont {C.~J.}\ \bibnamefont
  {Foot}},\ }\href@noop {} {\emph {\bibinfo {title} {Atomic Physics}}}\
  (\bibinfo  {publisher} {Oxford University Press},\ \bibinfo {address}
  {Oxford},\ \bibinfo {year} {2004})\BibitemShut {NoStop}%
\bibitem [{\citenamefont {Savard}\ \emph {et~al.}(1997)\citenamefont {Savard},
  \citenamefont {O'Hara},\ and\ \citenamefont {Thomas}}]{Savard1997}%
  \BibitemOpen
  \bibfield  {author} {\bibinfo {author} {\bibfnamefont {T.~A.}\ \bibnamefont
  {Savard}}, \bibinfo {author} {\bibfnamefont {K.~M.}\ \bibnamefont {O'Hara}},
  \ and\ \bibinfo {author} {\bibfnamefont {J.~E.}\ \bibnamefont {Thomas}},\
  }\bibfield  {title} {\enquote {\bibinfo {title} {Laser-noise-induced heating
  in far-off resonance optical traps},}\ }\href {\doibase 10.1103/PhysRevA.56.R1095} {\bibfield  {journal} {\bibinfo  {journal} {Phys.
  Rev. A}\ }\textbf {\bibinfo {volume} {56}},\ \bibinfo {pages} {R1095}
  (\bibinfo {year} {1997})}\BibitemShut {NoStop}%
\bibitem [{\citenamefont {Gehm}\ \emph {et~al.}(1998)\citenamefont {Gehm},
  \citenamefont {O'Hara}, \citenamefont {Savard},\ and\ \citenamefont
  {Thomas}}]{Gehm1998}%
  \BibitemOpen
  \bibfield  {author} {\bibinfo {author} {\bibfnamefont {M.~E.}\ \bibnamefont
  {Gehm}}, \bibinfo {author} {\bibfnamefont {K.~M.}\ \bibnamefont {O'Hara}},
  \bibinfo {author} {\bibfnamefont {T.~A.}\ \bibnamefont {Savard}}, \ and\
  \bibinfo {author} {\bibfnamefont {J.~E.}\ \bibnamefont {Thomas}},\ }\bibfield
   {title} {\enquote {\bibinfo {title} {Dynamics of noise-induced heating in
  atom traps},}\ }\href {\doibase 10.1103/PhysRevA.58.3914} {\bibfield
  {journal} {\bibinfo  {journal} {Phys. Rev. A}\ }\textbf {\bibinfo {volume}
  {58}},\ \bibinfo {pages} {3914} (\bibinfo {year} {1998})}\BibitemShut
  {NoStop}%
\bibitem [{\citenamefont {Tian}\ \emph {et~al.}(2019)\citenamefont {Tian},
  \citenamefont {Wang}, \citenamefont {Yang}, \citenamefont {Zhang},
  \citenamefont {Li},\ and\ \citenamefont {Zhang}}]{Tian2019}%
  \BibitemOpen
  \bibfield  {author} {\bibinfo {author} {\bibfnamefont {Y.-L.}\ \bibnamefont
  {Tian}}, \bibinfo {author} {\bibfnamefont {Z.-H.}\ \bibnamefont {Wang}},
  \bibinfo {author} {\bibfnamefont {P.-F.}\ \bibnamefont {Yang}}, \bibinfo
  {author} {\bibfnamefont {P.-F.}\ \bibnamefont {Zhang}}, \bibinfo {author}
  {\bibfnamefont {G.}~\bibnamefont {Li}}, \ and\ \bibinfo {author}
  {\bibfnamefont {T.-C.}\ \bibnamefont {Zhang}},\ }\bibfield  {title} {\enquote
  {\bibinfo {title} {Comparison of single-neutral-atom qubit between in bright
  trap and in dark trap},}\ }\href {\doibase 10.1088/1674-1056/28/2/023701}
  {\bibfield  {journal} {\bibinfo  {journal} {Chin. Phys. B}\ }\textbf
  {\bibinfo {volume} {28}},\ \bibinfo {pages} {023701} (\bibinfo {year}
  {2019})}\BibitemShut {NoStop}%
\bibitem [{\citenamefont {F\"orster}\ \emph {et~al.}(2009)\citenamefont
  {F\"orster}, \citenamefont {Karski}, \citenamefont {Choi}, \citenamefont
  {Steffen}, \citenamefont {Alt}, \citenamefont {Meschede}, \citenamefont
  {Widera}, \citenamefont {Montano}, \citenamefont {Lee}, \citenamefont
  {Rakreungdet},\ and\ \citenamefont {Jessen}}]{Forster2009}%
  \BibitemOpen
  \bibfield  {author} {\bibinfo {author} {\bibfnamefont {L.}~\bibnamefont
  {F\"orster}}, \bibinfo {author} {\bibfnamefont {M.}~\bibnamefont {Karski}},
  \bibinfo {author} {\bibfnamefont {J.-M.}\ \bibnamefont {Choi}}, \bibinfo
  {author} {\bibfnamefont {A.}~\bibnamefont {Steffen}}, \bibinfo {author}
  {\bibfnamefont {W.}~\bibnamefont {Alt}}, \bibinfo {author} {\bibfnamefont
  {D.}~\bibnamefont {Meschede}}, \bibinfo {author} {\bibfnamefont
  {A.}~\bibnamefont {Widera}}, \bibinfo {author} {\bibfnamefont
  {E.}~\bibnamefont {Montano}}, \bibinfo {author} {\bibfnamefont {J.~H.}\
  \bibnamefont {Lee}}, \bibinfo {author} {\bibfnamefont {W.}~\bibnamefont
  {Rakreungdet}}, \ and\ \bibinfo {author} {\bibfnamefont {P.~S.}\ \bibnamefont
  {Jessen}},\ }\bibfield  {title} {\enquote {\bibinfo {title} {Microwave
  control of atomic motion in optical lattices},}\ }\href {\doibase 10.1103/PhysRevLett.103.233001} {\bibfield  {journal} {\bibinfo  {journal}
  {Phys. Rev. Lett.}\ }\textbf {\bibinfo {volume} {103}},\ \bibinfo {pages}
  {233001} (\bibinfo {year} {2009})}\BibitemShut {NoStop}%
\bibitem [{\citenamefont {Li}\ \emph {et~al.}(2012)\citenamefont {Li},
  \citenamefont {Corcovilos}, \citenamefont {Wang},\ and\ \citenamefont
  {Weiss}}]{LiX2012}%
  \BibitemOpen
  \bibfield  {author} {\bibinfo {author} {\bibfnamefont {X.}~\bibnamefont
  {Li}}, \bibinfo {author} {\bibfnamefont {T.~A.}\ \bibnamefont {Corcovilos}},
  \bibinfo {author} {\bibfnamefont {Y.}~\bibnamefont {Wang}}, \ and\ \bibinfo
  {author} {\bibfnamefont {D.~S.}\ \bibnamefont {Weiss}},\ }\bibfield  {title}
  {\enquote {\bibinfo {title} {3d projection sideband cooling},}\ }\href
  {\doibase 10.1103/PhysRevLett.108.103001} {\bibfield  {journal} {\bibinfo
  {journal} {Phys. Rev. Lett.}\ }\textbf {\bibinfo {volume} {108}},\ \bibinfo
  {pages} {103001} (\bibinfo {year} {2012})}\BibitemShut {NoStop}%
\bibitem [{\citenamefont {Kaufman}\ \emph {et~al.}(2012)\citenamefont
  {Kaufman}, \citenamefont {Lester},\ and\ \citenamefont
  {Regal}}]{Kaufman2012}%
  \BibitemOpen
  \bibfield  {author} {\bibinfo {author} {\bibfnamefont {A.~M.}\ \bibnamefont
  {Kaufman}}, \bibinfo {author} {\bibfnamefont {B.~J.}\ \bibnamefont {Lester}},
  \ and\ \bibinfo {author} {\bibfnamefont {C.~A.}\ \bibnamefont {Regal}},\
  }\bibfield  {title} {\enquote {\bibinfo {title} {Cooling a single atom in an
  optical tweezer to its quantum ground state},}\ }\href {\doibase 10.1103/PhysRevX.2.041014} {\bibfield  {journal} {\bibinfo  {journal} {Phys.
  Rev. X}\ }\textbf {\bibinfo {volume} {2}},\ \bibinfo {pages} {041014}
  (\bibinfo {year} {2012})}\BibitemShut {NoStop}%
\bibitem [{\citenamefont {Thompson}\ \emph {et~al.}(2013)\citenamefont
  {Thompson}, \citenamefont {Tiecke}, \citenamefont {Zibrov}, \citenamefont
  {Vuleti\ifmmode~\acute{c}\else \'{c}\fi{}},\ and\ \citenamefont
  {Lukin}}]{Thompson2013}%
  \BibitemOpen
  \bibfield  {author} {\bibinfo {author} {\bibfnamefont {J.~D.}\ \bibnamefont
  {Thompson}}, \bibinfo {author} {\bibfnamefont {T.~G.}\ \bibnamefont
  {Tiecke}}, \bibinfo {author} {\bibfnamefont {A.~S.}\ \bibnamefont {Zibrov}},
  \bibinfo {author} {\bibfnamefont {V.}~\bibnamefont
  {Vuleti\ifmmode~\acute{c}\else \'{c}\fi{}}}, \ and\ \bibinfo {author}
  {\bibfnamefont {M.~D.}\ \bibnamefont {Lukin}},\ }\bibfield  {title} {\enquote
  {\bibinfo {title} {Coherence and raman sideband cooling of a single atom in
  an optical tweezer},}\ }\href {\doibase 10.1103/PhysRevLett.110.133001}
  {\bibfield  {journal} {\bibinfo  {journal} {Phys. Rev. Lett.}\ }\textbf
  {\bibinfo {volume} {110}},\ \bibinfo {pages} {133001} (\bibinfo {year}
  {2013})}\BibitemShut {NoStop}%
\bibitem [{\citenamefont {Wang}\ \emph {et~al.}(2016)\citenamefont {Wang},
  \citenamefont {Kumar}, \citenamefont {Wu},\ and\ \citenamefont
  {Weiss}}]{Wang2016}%
  \BibitemOpen
  \bibfield  {author} {\bibinfo {author} {\bibfnamefont {Y.}~\bibnamefont
  {Wang}}, \bibinfo {author} {\bibfnamefont {A.}~\bibnamefont {Kumar}},
  \bibinfo {author} {\bibfnamefont {T.-Y.}\ \bibnamefont {Wu}}, \ and\ \bibinfo
  {author} {\bibfnamefont {D.~S.}\ \bibnamefont {Weiss}},\ }\bibfield  {title}
  {\enquote {\bibinfo {title} {Single-qubit gates based on targeted phase
  shifts in a 3d neutral atom array},}\ }\href {\doibase 10.1126/science.aaf2581} {\bibfield  {journal} {\bibinfo  {journal}
  {Science}\ }\textbf {\bibinfo {volume} {352}},\ \bibinfo {pages} {1562}
  (\bibinfo {year} {2016})}\BibitemShut {NoStop}%
\bibitem [{\citenamefont {Barredo}\ \emph {et~al.}(2020)\citenamefont
  {Barredo}, \citenamefont {Lienhard}, \citenamefont {Scholl}, \citenamefont
  {de~L\'es\'eleuc}, \citenamefont {Boulier}, \citenamefont {Browaeys},\ and\
  \citenamefont {Lahaye}}]{Barredo2020}%
  \BibitemOpen
  \bibfield  {author} {\bibinfo {author} {\bibfnamefont {D.}~\bibnamefont
  {Barredo}}, \bibinfo {author} {\bibfnamefont {V.}~\bibnamefont {Lienhard}},
  \bibinfo {author} {\bibfnamefont {P.}~\bibnamefont {Scholl}}, \bibinfo
  {author} {\bibfnamefont {S.}~\bibnamefont {de~L\'es\'eleuc}}, \bibinfo
  {author} {\bibfnamefont {T.}~\bibnamefont {Boulier}}, \bibinfo {author}
  {\bibfnamefont {A.}~\bibnamefont {Browaeys}}, \ and\ \bibinfo {author}
  {\bibfnamefont {T.}~\bibnamefont {Lahaye}},\ }\bibfield  {title} {\enquote
  {\bibinfo {title} {Three-dimensional trapping of individual rydberg atoms in
  ponderomotive bottle beam traps},}\ }\href {\doibase 10.1103/PhysRevLett.124.023201} {\bibfield  {journal} {\bibinfo  {journal}
  {Phys. Rev. Lett.}\ }\textbf {\bibinfo {volume} {124}},\ \bibinfo {pages}
  {023201} (\bibinfo {year} {2020})}\BibitemShut {NoStop}%
\bibitem [{\citenamefont {Zhang}\ \emph {et~al.}(2011)\citenamefont {Zhang},
  \citenamefont {Robicheaux},\ and\ \citenamefont {Saffman}}]{Zhang2011}%
  \BibitemOpen
  \bibfield  {author} {\bibinfo {author} {\bibfnamefont {S.}~\bibnamefont
  {Zhang}}, \bibinfo {author} {\bibfnamefont {F.}~\bibnamefont {Robicheaux}}, \
  and\ \bibinfo {author} {\bibfnamefont {M.}~\bibnamefont {Saffman}},\
  }\bibfield  {title} {\enquote {\bibinfo {title} {Magic-wavelength optical
  traps for rydberg atoms},}\ }\href {\doibase 10.1103/PhysRevA.84.043408}
  {\bibfield  {journal} {\bibinfo  {journal} {Phys. Rev. A}\ }\textbf {\bibinfo
  {volume} {84}},\ \bibinfo {pages} {043408} (\bibinfo {year}
  {2011})}\BibitemShut {NoStop}%
\end{thebibliography}
\end{document}